\documentclass[a4paper,UKenglish]{lipics}
\usepackage{microtype}
\usepackage{amsmath}
\usepackage{amsthm}

\newcommand{\ignore}[1]{}

\theoremstyle{plain}
\newtheorem{rmk}[theorem]{Remark}
\newtheorem{proposition}[theorem]{Proposition}

\usepackage{times}
\usepackage{helvet}
\usepackage{courier}
\usepackage{color}
\usepackage{multicol}

\usepackage{amssymb}
\usepackage{graphicx}
\usepackage{color}
\usepackage{hhline}

\newcommand{\boxtheorem}{\hfill $\Box$\vspace{-2mm}}
\newcommand{\nit}[1]{{\it #1}}

\newcommand{\IC}{\nit{IC}}

\def\IC{{\textit{IC}}}
\usepackage{graphicx}

\usepackage{algpseudocode}
\usepackage{algorithm}

\newcommand{\red}[1]{\textcolor{red}{#1}}
\newcommand{\blue}[1]{\textcolor{blue}{#1}}

\newcommand{\comlb}[1]{{\vspace{2mm}\noindent \red{\bf COMM(LEO):}}~ #1 \hfill {\bf    END.}\\}
\newcommand{\combabak}[1]{{\vspace{4mm}\noindent \bf  COMM(Babak):}~ \red {\em  #1}\hfill {\bf END.}\\}

\newcommand{\defproof}[2]{{\noindent\bf Proof of #1:\
}#2 \boxtheorem\\}

\newcommand{\mc}[1]{\mathcal{ #1}}
\newcommand{\mf}[1]{\mathfrak{ #1}}

\ignore{
\volumeinfo{Marcelo Arenas and Martin Ugarte}
{2}
{18th International Conference on Database Theory (ICDT'15)}
{31}
{1}
{1}
\EventShortName{ICDT'15}
\DOI{10.4230/LIPIcs.ICDT.2015.1}
}

\title{\bf From Causes for Database Queries to Repairs and Model-Based Diagnosis and  Back}

\author[1]{Babak Salimi}
\author[2]{Leopoldo Bertossi}
\affil[1]{Carleton University,
  School of Computer Science,  Ottawa,  Canada\\
  \texttt{bsalimi@scs.carleton.ca}}
  \affil[2]{Carleton University,
  School of Computer Science,  Ottawa,  Canada\\
  \texttt{bertossi@scs.carleton.ca}}


\begin{document}
\maketitle

\begin{abstract}
In this work we establish and investigate connections between causality for query answers in databases, database repairs wrt. denial constraints, and consistency-based diagnosis.
The first two are relatively new problems in databases, and the third one is an established subject in knowledge representation. We show how to obtain database repairs from causes and
the other way around. Causality problems are formulated as diagnosis problems, and the diagnoses provide causes and their responsibilities. The vast
body of research on database repairs can be applied to the newer problem of determining actual causes for query answers and their responsibilities.
 These connections, which are interesting {\em per se}, allow us, after a transition -inspired by consistency-based diagnosis-
 to computational problems on hitting sets and vertex covers
 in hypergraphs,  to obtain several new algorithmic and complexity results for causality in databases.
\end{abstract}

\section{Introduction}
\vspace{-1mm}
When querying a database, a user may not always obtain the expected results, and the system could provide some explanations. They could be useful to further understand the data or  check if the
query is the intended one. Actually,
the notion of explanation for a query result was introduced in \cite{Meliou2010a}, on the basis of the deeper concept of {\em actual causation}.

A tuple $t$  is an {\em actual cause} for an answer $\bar{a}$ to a
conjunctive query $\mc{Q}$ from  a relational database instance $D$ if there is a {\em contingent} set of tuples $\Gamma$,
such that, after removing $\Gamma$ from $D$, $\bar{a}$ is still an answer, but after further  removing $t$ from $D\smallsetminus \Gamma$, $\bar{a}$ is not an answer anymore.
Here, $\Gamma$ is a set of tuples that has to accompany $\bar{a}$ for it to be a cause.
Actual causes and contingent tuples  are restricted to be among a pre-specified set
of {\em endogenous tuples}, which are admissible, possible candidates for causes, as opposed to {\em exogenous tuples}, which may also be present in the database. In rest of this paper, whenever we simply say ``cause", we mean ``actual cause".

In applications involving large
data sets, it is crucial to rank potential causes by their {\em responsibilities} \cite{Meliou2010b,Meliou2010a}, which reflect the relative (quantitative) degrees of their causality
for a query result. The responsibility measure for a cause is based on its {\em contingency sets}: the smallest (one of) its contingency sets, the strongest it is as a cause.

Actual causation, as used in  \cite{Meliou2010a}, can be traced back to
\cite{Halpern01,Halpern05}, which provides  a model-based account of causation on the basis of {\em counterfactual dependence}. Responsibility was introduced in \cite{cs-AI-0312038}, to capture the
intuitive notion of  {\em degree of causation}.

Apart from the explicit use of causality, research on explanations for query results has focused mainly, and rather implicitly, on provenance
\cite{BunemanKT01,BunemanT07,Cheney09,CuiWW00,Tannen10,Karvounarakis02,tannen}\ignore{, and
more recently, on provenance for non-answers \cite{ChapmanJ09,HuangCDN08}}.\ignore{\footnote{That is, tracing back, sometimes through the interplay of database tuple annotations, the reasons for {\em not} obtaining a possibly  expected answer to a query.}}
 A close connection between
causality and provenance has been established in \cite{Meliou2010a}.  However, causality is a more refined notion that identifies causes
for query results on the basis of  user-defined criteria, and ranks causes according to
their responsibilities \cite{Meliou2010b}. \ignore{For a formalization of non-causality-based explanations for
query answers in DL ontologies, see \cite{borgida}.  }

{\em Consistency-based diagnosis}  \cite{Reiter87}, a form of model-based diagnosis \cite[sec. 10.3]{struss}, is an area of knowledge representation. The problem here is, given the {\em specification} of a system in some logical formalism and a usually unexpected {\em observation}  about the system, to obtain {\em explanations} for the observation, in the form of a diagnosis for the unintended behavior.

In a different direction, a database instance, $D$, that is expected to satisfy certain integrity constraints may fail to do so. In this case, a {\em repair} of $D$ is a
database   $D'$ that does satisfy the integrity constraints and {\em minimally departs} from $D$. Different forms of minimality can be applied and investigated. A {\em consistent answer} to a query
from $D$ and wrt.\ the integrity constraints is a query answer that is obtained from all possible repairs, i.e. is invariant or certain under the class of repairs. These notions were introduced
in \cite{pods99} (surveys of the area can be found in \cite{Bertossi06,2011Bertossi}). Although not in the context of repairs, consistency-based diagnosis has been applied
to consistency restoration of a database wrt.  integrity constraints \cite{Gertz97}.

These three forms of reasoning, namely inferring causes from databases, consistency-based diagnosis, and consistent query answering (and repairs) are all {\em non-monotonic} \cite{Saliminmr2014}.
For example, a (most responsible) cause for a query result may not be such anymore after the database is updated. Furthermore, they all reflect some sort of
{\em uncertainty} about the information at hand. In this work we establish natural, precise, useful, and deeper connections
between these three reasoning tasks.

More precisely,  we unveil a
strong connection between computing causes and their responsibilities for conjunctive query answers, on one hand, and computing repairs in databases
 wrt. {\em denial constraints}, on the other. These computational
problems can be reduced to each other. In order to obtain repairs wrt. {\em a set} of denial constraints from causes, we investigate causes for queries
that are {\em unions of conjunctive queries}, and develop algorithms to compute  causes and responsibilities.

We show that inferring and computing actual causes  and their responsibilities in a database setting become
 diagnosis reasoning problems and tasks.  Actually,
a causality-based explanation for a conjunctive query answer can be viewed as a diagnosis, where in essence the first-order logical reconstruction of the relational database
 provides the system description  \cite{Reiter82}, and the observation is the query answer. We also establish  a bidirectional connection between diagnosis and
 repairs.

 Being the causality problems the main focus of this work, we take advantage of algorithms and complexity results both for consistency-based diagnosis; and
 database repairs and consistent query answering \cite{2011Bertossi}. In this way, we obtain new complexity results  for the main problems of causality, namely
 computing actual causes, determining their responsibilities, and obtaining most responsible causes; and also for their decision versions. In particular,
we obtain fixed-parameter tractable algorithms for some of them.
 More precisely, our main results  are as follows:\footnote{A few of the results included here appear in \cite{Saliminmr2014}.} \ (the complexity results are all  in data complexity)

 \vspace{1mm}
 \noindent 1. \ For a boolean conjunctive query and its associated denial constraint (the former being its violation view), we establish a precise connection (characterization and computational reductions)
 between actual causes for the query (being true) and the subset- and cardinality-repairs of the instance wrt.\ the denial constraint. We obtain causes from repairs.

 \noindent 2. \ We obtain repairs from causes, for which we extend the treatment of causality to unions of conjunctive queries  (to represent multiple denial constraints).  We characterize an actual cause's
 responsibility in terms of cardinality-repairs. We provide algorithms to compute causes and their (minimal) contingency sets for unions of conjunctive queries. The causes can be computed in PTIME.

 \noindent 3. \ We establish a precise connection between consistency-based diagnosis for a boolean conjunctive query being unexpectedly true according to a
 system description, and causes for the query being true. In particular, we show how to compute actual causes, their contingency sets, and responsibilities
 using the diagnosis characterization. Hitting-set-based algorithmic approaches to diagnosis inspire our algorithmic/complexity  approaches to causality.

 \noindent 4. \ We reformulate the causality problems as hitting set problems and  vertex cover
 problems on hypergraphs, which allows us to apply results and techniques for the latter to causality.

 \noindent 5. \ (a) Checking minimal contingency sets can be done in PTIME. \ (b) The responsibility (decision)  problem for conjunctive queries  becomes \nit{NP}-complete. \ (c) However, it is fixed-parameter tractable when the
 parameter is the inverse of the responsibility bound. \ (d)  The functional problem of computing the causes' responsibilities is $\nit{FP}^{\nit{NP(log(n)})}$-complete, and
 deciding most responsible causes is  $P^\nit{NP(log(n))}$-complete.

 \noindent 6. \ The structure of the resulting hitting-set problem allows us to obtain efficient
 parameterized algorithms and
 good approximation algorithms for computing causes and minimal contingency sets.

  \noindent 7. \ On the basis of the causality/repair connection, and the dichotomy result for causality \cite{Meliou2010a}, we obtain a dichotomy result for the complexity
  of deciding the existence of repairs of a certain size wrt. single, self-join-free denial constraints.

 \noindent 8. \ We discuss extensions and open issues that deserve investigation.

\vspace{2mm}
\noindent The paper is structured as follows. Section \ref{sec:prel} introduces technical preliminaries for relational databases, causality in databases, database repairs and consistent query answering, consistency-based
 diagnosis, and relevant complexity classes. Section \ref{sec:causfrepair}  characterizes actual causes and responsibilities in terms of database repairs.
Section \ref{sec:causfrepair} characterizes repairs and consistent query answering in terms of causes and contingency sets for queries that are unions of conjunctive queries; and presents an algorithm
for computing both of the latter. Section \ref{sec:MBDtoRep} formulates causality problems as consistency-based diagnosis problems, and the latter as
repair problems. Section \ref{sec:MBDcomx} shows complexity and algorithmic results; in particular a fixed-parameter tractability result for causes' responsibilities.
Finally, Section \ref{sec:disc} discusses several relevant issues, connections and open problems around causality in databases. Proofs of results without an implicit proof in the main body of this paper can be found in
the appendix.

\vspace{-1mm}
\section{Preliminaries}\label{sec:prel}
\vspace{-1mm}We consider relational database schemas of the form $\mathcal{S} = (U,\mc{P})$,  where $U$ is the possibly infinite
database domain of {\em constants} and $\mc{P}$ is a finite set of {\em database predicates}\footnote{As opposed to built-in predicates (e.g. $\neq$) that we assume
do not appear, unless explicitly stated otherwise.} of fixed arities. A database instance $D$
compatible with $\mathcal{S}$ can be seen as a finite set of ground atomic formulas (in databases aka. atoms or tuples), of the form $P(c_1, ..., c_n)$, where $P \in \mc{P}$ has arity $n$, and $c_1, \ldots , c_n \in U$.
A {\em conjunctive query}  (CQ) is a formula $\mc{Q}(\bar{x})$ of the first-order (FO) logic language, $\mc{L}(\mc{S})$, associated to $\mc{S}$ of the form \ $\exists \bar{y}(P_1(\bar{t}_1) \wedge \cdots \wedge P_m(\bar{t}_m))$,
where the $P_i(\bar{t}_i)$ are atomic formulas, i.e. $P_i \in \mc{P}$, and the $\bar{t}_i$ are sequences of terms, i.e. variables or constants. The $\bar{x}$ in  $\mc{Q}(\bar{x})$ shows
all the free variables in the formula, i.e. those not appearing in $\bar{y}$. If $\bar{x}$ is non-empty, the query is {\em open}. If $\bar{x}$ is empty, the query is {\em boolean} (a BCQ), i.e. the query is a sentence, in which case, it is true or false
in a database, denoted by $D \models \mc{Q}$ and $D \not\models \mc{Q}$, respectively. A sequence $\bar{c}$ of constants is an answer to an open query $\mc{Q}(\bar{x})$ if $D \models \mc{Q}[\bar{c}]$, i.e.
the query becomes true in $D$ when the variables are replaced by the corresponding constants in $\bar{c}$.

An {\em integrity constraint} is a sentence of language $\mc{L}(\mc{S})$, and then, may be true or false
in an instance for schema $\mc{S}$. Given a set $\IC$ of integrity constraints, a database instance $D$ is {\em consistent} if $D \models \IC$; otherwise it is said to be {\em inconsistent}.
In this work we assume that sets of integrity constraints are always finite and logically consistent.
A particular class of  integrity constraints  is formed by {\em denial constraints} (DCs), which are sentences  $\kappa$ of the form:
 $\forall \bar{x} \neg  (A_1(\bar{x}_1)  \wedge \cdots  \wedge A_n(\bar{x}_n)$, where $\bar{x}= \bigcup \bar{x}_i$ and each $A_i(\bar{x}_i)$ is a database atom, i.e. predicate $A \in \mc{P}\!$.  (The atoms may contain
 constants.)
Denial constraints  are  exactly the negations of BCQs.

\vspace{2mm}
\noindent {\bf Causality and Responsibility.} \ \  Assume that the database instance is split in two, i.e. $D=D^n \cup D^x$,  where $D^n$ and $D^x$ denote the sets of {\em endogenous} and {\em exogenous} tuples,
 respectively.
A tuple $t \in D^n$ is called a
{\em counterfactual cause} for  a BCQ $\mc{Q}$,  if $D\models \mc{Q}$ and $D\smallsetminus \{t\}  \not \models \mc{Q}$.  A tuple $t \in D^n$ is an {\em actual cause} for  $\mc{Q}$
if there  exists $\Gamma \subseteq D^n$, called a {\em contingency set},  such that $t$ is a counterfactual cause for $\mc{Q}$ in $D\smallsetminus \Gamma$ \ \cite{Meliou2010a}. \
We will concentrate mostly on CQs. However, the definition of actual causes and contingency sets  can be applied without a change
to monotone queries in general \cite{Meliou2010a}.

 The {\em responsibility} of an actual cause $t$ for $\mc{Q}$, denoted by $\rho_{_{\!D\!}}(t)$,  is the numerical value $\frac{1}{|\Gamma| + 1}$, where $|\Gamma|$ is the
size of the smallest contingency set for $t$. We can extend responsibility to all the other tuples in $D^n$ by setting their value to $0$. Those tuples are not actual causes for $\mc{Q}$.

\ignore{
In \cite{Meliou2010a}, causality for non-query answers is defined on basis of sets of {\em potentially missing tuples} that account for the missing answer. Causality for
non-answers becomes a variation of causality for answers. In this work we do not consider non-query answers.}

\begin{example}\label{ex:cfex1}
Consider $D = D^n = \{R(a_4,a_3), R(a_2,a_1), R(a_3,a_3), S(a_4),$ $S(a_2),$ $S(a_3)\}$, and the query $\mc{Q} :\exists x \exists y ( S(x) \land R(x, y) \land S(y))$. It holds: $D \models \mc{Q}$.

 Tuple $S(a_3)$ is a counterfactual cause for $\mc{Q}$. If $S(a_3)$ is removed from $D$,  $\mc{Q}$ is not true anymore. Therefore, the responsibility of $S(a_3)$ is 1. Besides, $R(a_4,a_3)$ is an actual cause for $\mc{Q}$ with contingency set
$\{ R(a_3,a_3)\}$. If $R(a_3,a_3)$ is removed from $D$, $\mc{Q}$ is still true, but further removing $R(a_4,a_3)$ makes $\mc{Q}$ false. The responsibility of $R(a_4,a_3)$ is $\frac{1}{2}$, because its smallest contingency sets have size $1$. Likewise,  $R(a_3,a_3)$ and $S(a_4)$ are actual causes for $\mc{Q}$ with responsibility  $\frac{1}{2}$.

For the  same $\mc{Q}$, but with
$D=\{S(a_3),S(a_4),R(a_4,a_3) \}$, and the
partition $D^n=\{S(a_4),S(a_3)\}$ and $D^x=\{ R(a_4,a_3)\}$, it turns out that both $S(a_3)$
and $S(a_4)$ are counterfactual
causes for $\mc{Q}$.\boxtheorem
\end{example}

\noindent {\bf \em Notation:} \ $\mc{CS}(D^n,D^x,\mc{Q})$ denotes the set of actual causes for BCQ  $\mc{Q}$ (being true) from instance $D=D^n \cup D^x$. When
$D^n = D$ and $D^x = \emptyset$, we sometimes simply write: $\mc{CS}(D,\mc{Q})$.

\vspace{2mm}
\noindent {\bf Database Repairs.} \ \
Given a set $\IC$ of integrity constraints,  a {\em subset repair} (simply, S-repair) of a possibly inconsistent instance $D$ for schema $\mc{S}$  is an instance $D'$ for
$\mc{S}$ that satisfies $\IC$ and makes $\Delta(D,D')=(D \smallsetminus D') \cup( D' \smallsetminus D)$ minimal under set inclusion.
$\nit{Srep}(D,\IC)$ denotes the set of S-repairs of $D$ wrt.\ $\IC$ \cite{pods99}.
Similarly, $D'$ is a  {\em cardinality repair} (simply C-repair) of $D$ if $D'$ satisfies $\IC$ and minimizes $|\Delta(D,D')|$.
$\nit{Crep}(D,\IC)$ denotes the class of C-repairs of $D$ wrt.\ $\IC$.
C-repairs are S-repairs of minimum cardinality.

For DCs, S-repairs and C-repairs are obtained from the original instance by deleting an S-minimal, resp. C-minimal, set of tuples.\footnote{We will usually say that a set is S-minimal in a class of sets $\mc{C}$ if it minimal under set inclusion in $\mc{C}$.
Similarly, a set is C-minimal if it is minimal in cardinality within $\mc{C}$.}  More generally, different {\em repair semantics} may be considered to restore consistency wrt. general integrity constraints. They depend on the kind of allowed updates on the database (i.e.
tuple insertions/deletions, changes of attribute values), and the minimality
conditions on repairs (e.g. subset-minimality, cardinality-minimality, etc.). Given $D$ and $\IC$, a repair semantics determines a class of intended or preferred repairs \cite[sec. 2.5]{2011Bertossi}.

Given a repair semantics, \nit{RS}, \ $\bar{c}$ is a {\em consistent answer} to an open query $\mc{Q}(\bar{x})$ if $D' \models \mc{Q}[\bar{c}]$ for every \nit{RS}-repair $D'\!$.  A BCQ is {\em consistently true} if it is true in all \nit{RS}-repairs. If $\bar{c}$ is a consistent answer to $\mc{Q}(\bar{x})$ wrt. S-repairs, we say
 it is an {\em S-consistent answer}. Similarly for {\em C-consistent answers}.\ignore{, \red{is denoted by $D \models_S \mc{Q}[\bar{c}]$,
resp. $D \models_C \mc{Q}[\bar{c}]$}.} Consistent query answering for DCs under S-repairs was investigated in detail \cite{Chomicki05}.  C-repairs and consistent query answering were investigated in detail in \cite{icdt07}.
(Cf. \cite{2011Bertossi} for more references.)

\vspace{2mm}
\noindent {\bf Consistency-Based Diagnosis.} \
Consistency-based diagnosis, a form of model-based diagnosis \cite[sec. 10.4]{struss}, considers problems $\mc{M}=(\nit{SD}, \nit{COMPS}, $ $ \nit{OBS})$, where $\nit{SD}$ is
the description in logic of the intended properties of a system under the {\em explicit} assumption that all the {\em components} in $\nit{COMPS}$,  are  working normally.
$\nit{OBS}$ is a FO sentence that represents the  observations.
If the system does not behave as expected (as shown by the observations),
then the logical theory obtained from $\nit{SD} \cup \nit{OBS}$ plus the explicit assumption, say $\bigwedge_{c \in \nit{COMPS}} \neg \nit{Ab}(c)$, that the components are indeed behaving normally, becomes inconsistent.
$\nit{Ab}$ is an {\em abnormality} predicate.\footnote{Here, and as usual, the atom $\nit{Ab}(c)$ expresses that component $c$ is (behaving) abnormal(ly).}

The inconsistency is captured via the {\em minimal conflict sets}, i.e. those minimal subsets $\nit{COMPS'}$ of $\nit{COMPS}$, such that $\nit{SD} \cup \nit{OBS} \cup \{\bigwedge_{c \in \nit{COMPS'}}
\neg \nit{Ab}(c)\}$ is  inconsistent. As expected, different notions of minimality can be used at this point. \ignore{It is common to use the distinguished predicate $\nit{Ab}(\cdot)$ for denoting {\em  abnormal}
(or abnormality). So, $\nit{Ab}(c)$ says that component $c$ is abnormal.}

 A {\em  minimal diagnosis} for $\mc{M}$ is a minimal subset $\Delta$ of $\nit{COMPS}$, such that $ \nit{SD} \cup \nit{OBS} \cup \{\neg \nit{Ab}(c)~|~c \in  \nit{COMPS} \smallsetminus \Delta \} \cup \{\nit{Ab}(c)~|~c \in \Delta\}$ is consistent. That is, consistency is restored by flipping the normality assumption to abnormality for a minimal set of components, and  those are the ones considered to be (jointly) faulty. The notion of minimality commonly used is S-minimality, i.e. a diagnosis that does  not have a proper subset that is a diagnosis. We will use this kind of minimality in relation to diagnosis. Diagnosis can be obtained from conflict sets \cite{Reiter87}.

\ignore{Diagnostic  reasoning  is  non-monotonic  in  the sense  that  a  diagnosis may not survive after the addition of  new
observations \cite{Reiter87}.  }

\ignore{\vspace{-3mm}
\paragraph{\bf Abductive diagnosis.}  In abductive diagnosis, we have a system description $\nit{SD}$ in a certain logical formalism, very commonly
in the form of a (extended) Datalog program \cite{EiterGL97}; a set $\nit{ABD}$ of atoms, called {\em abductible atoms} (or simply, {\em abductibles}), e.g. abnormality atoms of the form $\nit{Ab}(\bar{c})$); and an observation $\nit{OBS}$ as in
consistency-based diagnosis. The abduction problem is about computing a minimal (under certain minimality criterion) $\Delta \subseteq \nit{ABD}$ such that $\nit{SD} \cup \Delta \models \nit{OBS}$. For relationships
and comparisons between consistency-based and abductive diagnosis, see \cite{Console91}.  }

\vspace{2mm}
\noindent {\bf Complexity Classes.} \ \
We  recall some
complexity classes \cite{papa} used in this paper. $\nit{FP}$ is the class of functional problems
associated to decision problem in the class \nit{PTIME}, i.e. that are solvable
in polynomial time. $P^\nit{NP}$ (or $\Delta^P_2$) is the class of decision problems solvable in
polynomial time by a machine that makes calls to an $\nit{NP}$ oracle. For $\nit{P}^{\nit{NP(log} (n))}$
 the number of calls is logarithmic. It is not known if
$\nit{P}^{\nit{NP(log} (n))}$  is strictly contained in $\nit{P}^{\nit{NP}}\!\!$. \ $\nit{FP}^{\nit{NP(log} (n))}$ is
similarly defined.

\section{Actual Causes From Database Repairs} \label{sec:causfrepair}
\vspace{-1mm}
Let $D = D^n\cup D^x$ be an instance for schema $\mathcal{S}$, and
$\mc{Q}\!: \exists \bar{x}(P_1(\bar{x}_1) \wedge \cdots \wedge P_m(\bar{x}_m))$  a BCQ.
$\mc{Q}$ may be unexpectedly true, i.e.  $D \models \mc{Q}$.
Now, $\neg \mc{Q}$ is logically equivalent to the DC
$\kappa(\mc{Q})\!: \forall \bar{x} \neg (P_1(\bar{x}_1) \wedge \cdots \wedge P_m(\bar{x}_m))$.
 The requirement
that $\neg \mc{Q}$ holds can be captured by imposing $\kappa(\mc{Q})$ on $D$.
Due to $D \models \mc{Q}$, it holds $D \not \models \kappa(\mc{Q})$. So, $D$ is inconsistent wrt.\ $\kappa(\mc{Q})$, and could be repaired.

Repairs for (violations of) DCs are obtained
by tuple deletions.
Intuitively,  a tuple that participates in a violation of $\kappa(\mc{Q})$ in $D$ is an actual cause for $\mc{Q}$. S-minimal sets of tuples like this
are expected to correspond to S-repairs for $D$ and $\kappa(\mc{Q})$. More precisely,
given an instance $D = D^n \cup D^x$, a BCQ $\mc{Q}$, and a tuple $t \in D^n$, we consider the class containing the sets of differences between $D$ and those S- or C-repairs that do not contain $t$, and are obtained
by removing a subset of $D^n$: \vspace{-1mm}
\begin{eqnarray}
\hspace*{-2mm}\mc{DF}^s(D, D^n,\kappa(\mc{Q}), t)\!&=&\!\{ D \smallsetminus D'~|~ D' \in \nit{Srep}(D,\kappa(\mc{Q})),~ t \in (D\smallsetminus D') \subseteq D^n\},~~   \label{eq:df}\\
\hspace*{-2mm}\mc{DF}^c(D, D^n,\kappa(\mc{Q}), t)\!&=&\!\{ D \smallsetminus D'~|~ D' \in \nit{Crep}(D,\kappa(\mc{Q})),~ t \in (D\smallsetminus D') \subseteq D^n\}.~~  \label{eq:dfc}
\end{eqnarray}
\phantom{poto}

\vspace{-7mm}\noindent It holds $\mc{DF}^c(D, D^n,\kappa(\mc{Q}), t) \subseteq \mc{DF}^s(D, D^n,\kappa(\mc{Q}), t)$.
Now, any $s \in \mc{DF}^s(D, D^n,$ $\kappa(\mc{Q}), t)$ can be written as  $s=s' \cup \{t\}$.
From the S-minimality of S-repairs,    $D \smallsetminus (s' \cup \{t\}) \models \kappa(\mc{Q})$, but $D \smallsetminus s' \models \neg \kappa(\mc{Q})$, i.e.
$D \smallsetminus (s' \cup \{t\}) \not \models \mc{Q}$, but $D \smallsetminus s' \models \mc{Q}$. So,
$t$ is an actual cause for $\mc{Q}$ with contingency set $s'$. 

\vspace{-1mm}
\begin{proposition}\label{pro:c&r} \em
Given $D= D^n \cup D^x$, and a BCQ  $\mc{Q}$, $t \in D^n$ is an actual cause for $\mc{Q}$ \ iff \
$\mc{DF}^s(D, D^n,\kappa(\mc{Q}), t) \not = \emptyset$.\boxtheorem
\end{proposition}
\begin{proposition}\label{pro:r&r} \em
Given  $D= D^n \cup D^x$, a BCQ  $\mc{Q}$, and $t \in D^n$: \ (a)
If $\mc{DF}^s(D, D^n,$  $\kappa(\mc{Q}),  t) = \emptyset$, then $\rho(t)=0$. \ (b) Otherwise, $\rho(t)=\frac{1}{|s|}$, where $s \in \mc{DF}^s(D,D^n,$ $\kappa(\mc{Q}), t)$
and there is no $s' \in \mc{DF}^s(D, D^n,\kappa(\mc{Q}), t)$ such that $|s'| < |s|$. \boxtheorem
\end{proposition}
\begin{corollary}\label{cor:card} \em
Given  $D= D^n \cup D^x$, and a BCQ  $\mc{Q}$,  $t \in D^n$ is a most responsible actual cause for $\mc{Q}$ \ iff \
$\mc{DF}^c\!(D, D^n,\kappa(\mc{Q}), t) \not = \emptyset $.\boxtheorem
\end{corollary}
\begin{example}\label{ex:CausASrepex1} (ex. \ref{ex:cfex1} cont.) Consider the same instance $D$ and query $\mc{Q}$. In this case, the DC $\kappa(\mc{Q})$ is,
in Datalog notation, a  negative rule:  \ $\leftarrow S(x),R(x, y),S(y)$.

Here, $\nit{Srep}(D, \kappa(\mc{Q}))$ $=$ $\{D_1,$ $D_2,$ $D_3\}$ and $\nit{Crep}(D, \kappa(\mc{Q}))=\{D_1\}$, with
$D_1= \{R(a_4,a_3),$ $ R(a_2,a_1),$ $ R(a_3,a_3),$ $  S(a_4),$ $ S(a_2)\}$, \ $D_2 = \{ R(a_2,a_1),$ $ S(a_4),$
$S(a_2),$ $S(a_3)\}$, \ $D_3 = \{R(a_4,a_3),$ $ R(a_2,a_1),$ $ S(a_2),$ $ S(a_3)\}$.

For tuple $R(a_4,a_3)$, $\mc{DF}^s(D, D, \kappa(\mc{Q}), R(a_4,a_3))=\{D \smallsetminus D_2\} = \{ \{ R(a_4,a_3),$ $R(a_3 $ $,a_3)\} \}$, which, by Propositions  \ref{pro:c&r} and \ref{pro:r&r},
confirms that $R(a_4,a_3)$ is an actual cause,  with responsibility $\frac{1}{2}$.
\ For tuple $S(a_3)$,  $\mc{DF}^s(D, D, \kappa(\mc{Q}), $ $ S(a_3)) = \{D \smallsetminus D_1\}$ $=\{ S(a_3) \}$. So, $S(a_3)$
is an actual cause  with responsibility 1. Similarly, $R(a_3,a_3)$ is an actual cause with responsibility $\frac{1}{2}$, because  $\mc{DF}^s(D, D,\kappa(\mc{Q}),
R(a_3,a_3)) = \{D\smallsetminus D_2, \ D \smallsetminus D_3\}$ $=\{ \{R(a_4,$ $a_3),$ $R(a_3,a_3)\},$ $ \{R(a_3,a_3), S(a_4)\} \}$.

It holds  $\mc{DF}^s(D,$ $ D, \kappa(\mc{Q}),S(a_2)) = \mc{DF}^s(D, D, \kappa(\mc{Q}),R(a_2 , a_1)) = \emptyset$, because all repairs contain $S(a_2)$, $R(a_2 , a_1)$.
This means they do not participate in the violation of $\kappa(\mc{Q})$ or contribute to make $\mc{Q}$ true.
 So, they are not actual causes for $\mc{Q}$, confirming the result in Example \ref{ex:cfex1}.

$\mc{DF}^c(D, D, \kappa(\mc{Q}), S(a_3))=\{S(a_3)\}$. From Corollary \ref{cor:card},
$S(a_3)$  is the most responsible cause. \boxtheorem
\end{example}

\begin{rmk} \label{rem:ucq} \em
The results in this section can be easily extended to unions of BCQs without built-ins, i.e. essentially FO monotone queries without built-ins. This can be done by associating a DC to each disjunct of the query, and considering the corresponding problems for database repairs wrt. several DCs (cf. Section \ref{sec:disjcauses}).
\boxtheorem
\end{rmk}

\section{Database Repairs From Actual Causes} \label{sec:repairfcauses}
\vspace{-1mm}
We now characterize repairs for inconsistent databases wrt.\ {\em a
set of} DCs in terms of actual causes, and reduce their computation to computation of causes.
\ Consider an instance $D$  for  schema $\mathcal{S}$, and a set of
DCs $\Sigma$ on $\mc{S}$.  For each $\kappa \in \Sigma$, of the form \
 $\kappa\!: \ \leftarrow A_1(\bar{x}_1),\ldots,A_n(\bar{x}_n)$, consider its associated  {\em violation view} defined by a BCQ, namely \
 $V^{\!\kappa}\!\!: \
\exists\bar{x}(A_1(\bar{x}_1)\wedge \cdots \wedge A_n(\bar{x}_n))$.
Next, consider the query obtained as
the union of the individual violation views:  \
$V^{\Sigma}:= \bigvee_{\kappa \in  \Sigma} V^\kappa $, a {\em union of} BCQs (UBCQs). Clearly, $D$
violates (is inconsistent wrt.)  $\Sigma$ iff $D \models V^\Sigma\!\!$.
 It is easy to verify that $D$ is consistent wrt.\ $\Sigma$ iff
$\mc{CS}(D, \emptyset, V^\Sigma) = \emptyset$, i.e. there are no actual causes for $V^\Sigma$ to be true
when all tuples are endogenous.

Now, let us collect all
{\em S-minimal contingency sets} associated with an actual cause $t$ for
$V^\Sigma$: \vspace{-1mm}
\begin{eqnarray}
\mc{CT}(D,D^n,V^\Sigma,t) &:=& \{  s\subseteq D^n~|~D\smallsetminus s
\models V^\Sigma, ~D\smallsetminus (s \cup \{t\}) \not \models V^\Sigma,   \mbox{ and }
\label{eq:ct}\\
   && ~~~~~~~~~~~~~~~~~~~ \forall s''\subsetneqq s, \ D \smallsetminus (s'' \cup \{t\})
\models V^\Sigma \}. \nonumber
\end{eqnarray}
\phantom{poto}

\vspace{-6mm}\noindent
Notice that for  $s \in \mc{CT}(D,D^n,V^\Sigma,t)$, $t \notin s$.
If $t \in \mc{CS}(D, \emptyset, V^\Sigma)$ and $s \in \mc{CT}(D,D^n,V^\Sigma,t)$,  from  the
definition of actual cause and the S-minimality of
$s$, its holds that $s''= s \cup \{t\}$ is an S-minimal subset of $D$
with $D \smallsetminus s'' \not \models V^\Sigma$. So,  $D
\smallsetminus s''$
is an S-repair for $D$.  Then, the following holds.
\begin{proposition}\label{pro:sr&cp} \em For an instance $D$ and a set DCs
$\Sigma$,
$D' \subseteq D$ is an S-repair for $D$ wrt.\ $\Sigma$ iff,  for every $t
\in D \smallsetminus D'$:
$t \in \mc{CS}(D, \emptyset, V^\Sigma)$ and $D \smallsetminus (D' \cup
\{t\}) \in \mc{CT}(D, D,V^\Sigma, t)$.
\boxtheorem
\end{proposition}
To establish a connection between most responsible actual causes and
C-repairs, collect the most responsible actual causes for
$V^\Sigma$: \vspace{-1mm}
\begin{eqnarray*}
\mc{MRC}(D,V^\Sigma)&\!\!:=\!\!& \{t \in D~|~ t \in
\mc{CS}(D,\emptyset,V^\Sigma),  \not \exists t' \in
\mc{CS}(D,\emptyset,V^\Sigma) \mbox{ with } \rho(t')> \rho(t)   \}.
\end{eqnarray*}
\phantom{poto}

\vspace{-8mm}
\begin{proposition}\label{pro:cr&mrp} \em
For instance $D$ and   set of DCs $\Sigma$, $D' \subseteq D$ is a
C-repair for $D$ wrt.\ $\Sigma$ iff,  for every $t \in D \smallsetminus D'$:
$t \in  \mc{MRC}(D, V^\Sigma)$ and $D \smallsetminus (D' \cup \{t\}) \in
\mc{CT}(D, D,V^\Sigma, t)$.
\boxtheorem
\end{proposition}
Actual causes for $ V^\Sigma$,  with their contingency
sets, account for the violation of some $\kappa \in \Sigma$. Removing those tuples from
$D$ should remove the inconsistency. From Propositions \ref{pro:sr&cp} and \ref{pro:cr&mrp} we obtain: \vspace{-1mm}
\begin{corollary}\label{col:sr&cp} \em
Given an instance $D$ and a set DCs
$\Sigma$,  the instance obtained from $D$ by
removing an actual cause, resp.  a most responsible actual cause, for $ V^\Sigma$ together with any of its S-minimal, resp. C-minimal, contingency
sets forms an S-repair, resp. a C-repair, for $D$ wrt. $\Sigma$.
\boxtheorem
\end{corollary}
\begin{example}\label{ex:rc2cp}
 Consider $D=\{P(a),P(e),Q(a,b),R(a,c)\}$ and $\Sigma=\{\kappa_1,\kappa_2\}$, with
$\kappa_1\!: \ \leftarrow P(x), Q(x,y)$ and $\kappa_2 \!: \ \leftarrow
P(x), R(x,y)$. The violation views  are $ V^{\kappa_1}\!: \exists xy (P(x) \land Q(x,y)) $ and
 $ V^{\kappa_2} \!: \exists xy (P(x) \land R(x,y))$.
For $V^\Sigma :=  V^{\kappa_1}\lor V^{\kappa_2}$,
$D \models V^\Sigma$. $D$ is inconsistent wrt.\ $\Sigma$.

With all tuples endogenous,  $\mc{CS}(D, \emptyset, V^\Sigma)=\{P(a),Q(a,b),R(a,c) \}$. Its
elements are  associated with sets of S-minimal contingency sets: \
$\mc{CT}(D, D,V^\Sigma\!,Q(a,b))=\{\{ R(a,c)\}\}$, \ \ \
$\mc{CT}(D, D,$ $V^\Sigma\!, R(a,c))=\{ \{Q(a,b)\}\}$, \
$\mc{CT}(D,D,V^\Sigma\!,P(a))=\{\emptyset\}$.
From Corollary \ref{col:sr&cp},  and $\mc{CT}(D, D, V^\Sigma\!,$ $R(a,c))$,   $D_1=D \smallsetminus (\{
R(a,c)\} \cup \{Q(a,b)\}) =\{ P(a),P(e)\}$  is an S-repair. So is
$D_2=D \smallsetminus (\{P(a) \} \cup \emptyset)=\{P(e),Q(a,b),$ $R(a,c)\}$. These are the only S-repairs.

Furthermore, $\mc{MRC}(D, V^\Sigma)= \{P(a)\}$.  From
Corollary \ref{col:sr&cp}, $D_2$ is also a C-repair for $D$.
\boxtheorem
\end{example}
An actual cause $t$ with any of its S-minimal contingency sets determines a unique S-repair. The last example shows that, with different combinations of a cause and one of its contingency sets, we may obtain the same repair
(e.g. for the first two $\mc{CT}$s).
 So, we may have more minimal contingency sets than minimal repairs. However,   we may still have exponentially many minimal
 contingency sets, so as we may have exponentially many minimal repairs.

\vspace{-1.5mm}
\begin{example}\label{ex:repsVSconts}
Consider $D = \{R(1,0), R(1,1), \ldots, R(n,0), R(n,1), S(1), S(0)\}$ and the DC $\kappa\!: \ \leftarrow R(x,y), R(x,z), S(y),S(z)$. $D$ is inconsistent wrt. $\kappa$. There are exponentially many S-repairs of $D$: \
$D' = D \smallsetminus \{S(0)\}$,
 $D'' = D \smallsetminus \{S(1)\}$, $D_1 = D \smallsetminus \{R(1,0), \ldots, R(n,0)\}$, ..., $D_{2^n} = D \smallsetminus \{R(1,1), \ldots, R(n,1)\}$.
 The C-repairs are only $D'$ and $D''$.

For the BCQ $V^\kappa$ associated to $\kappa$, $D \models V^\kappa$, and $S(1)$ and $S(0)$ are actual causes for
$V^\kappa$ (courterfactual causes with responsibility $1$). All tuples in $R$ are  actual causes, each with exponentially many S-minimal contingency sets. For example,
$R(1, 0)$ has the S-minimal contingency set $\{R(2,0), \ldots, R(n,0)\}$, among  exponentially many others (any set
built with just one element from each of the pairs $\{R(2,0),R(2,1)\}$, ...,  $\{R(n,0), R(n,1)\}$  is one). \ignore{Therefore, $R(1, 0)$ has $2^{n-1}$
contingency sets. A similar argument holds for the rest of tuples in $R$.} \boxtheorem
\end{example}
The characterization results obtained so far extend those in  \cite{Saliminmr2014} for single DCs.

\vspace{-1mm}
\subsection{Causes for unions of conjunctive queries}\label{sec:disjcauses}

If we want to compute repairs wrt. sets of DCs from causes for UBCQs using, say Corollary \ref{col:sr&cp}, we first
need an algorithm for computing the actual causes
and their (minimal) contingency sets for  UBCQs.
These algorithms could be used as a first stage
for the  computation of S-repairs and C-repairs wrt. sets of DCs. However, these algorithms (cf. Section
\ref{sec:disjcausesCont}) are also interesting {\em per se}.

The PTIME algorithm for computing actual causes in \cite{Meliou2010a} is for single conjunctive queries, but does not
compute the actual causes' contingency sets.
 Actually, doing the latter increases the complexity, because {\em deciding responsibility}\footnote{For a precise formulation, see
 Definition \ref{def:resp}.} of actual causes is $\nit{N\!P}$-hard \cite{Meliou2010a} (which would be tractable
 if we could efficiently compute all (minimal) contingency sets).\footnote{Actually, \cite{Meliou2010a} presents a PTIME algorithm
for computing responsibilities  for a restricted class of CQs.} In principle, an algorithm for responsibilities can be used to compute C-minimal contingency sets, by iterating over all candidates,
but Example \ref{ex:repsVSconts} shows that there  can be exponentially many of them.

We first concentrate on the problem of computing actual causes for UBCQs, without their contingency sets, which requires some notation. \vspace{-1.5mm}
\begin{definition} \label{def:hsStuff}  Given  $\mc{Q} =  C_1 \vee \cdots \vee C_k$, with each $C_i$ a BCQ, and an instance $D$, \ (a) $\mathfrak{S}(D)$ is the collection of all
 S-minimal subsets of $D$ that satisfy a
disjunct  $C_i $ of $\mc{Q}$. \ (b) $\mf{S}^n(D)$ consists of the S-minimal subsets $s$ of  $D^n$ for which there exists a  $s' \! \in \mf{S}(D) $ with
$s \subseteq s'$ and $ s \smallsetminus s' \subseteq D^x$. \boxtheorem
\end{definition}
$\mf{S}^n(D)$  contains all S-minimal sets of endogenous tuples that simultaneously (and possibly accompanied by exogenous tuples) make the query true.
It is easy to see that  $\mathfrak{S}(D)$ and $\mf{S}^n(D)$ can be computed in polynomial time in the size of $D$.
Now, generalizing a result for CQs in \cite{Meliou2010a}, actual causes for a UBCQs can be computed in PTIME in the size of $D$ without computing contingency
 sets.
\begin{proposition}\label{pro:UBCQCausesindirect} \em
Given $D=D^x \cup D^n$ and  a UBCQ $\mc{Q}$, \ (a) \ $t$ is an actual cause for $\mc{Q}$ iff  there is ~ $s \! \in  \! \mf{S}^n(D)$ with $t \in s$. \
(b) The decision problem (about membership of)
$\nit{CPD} := \{ (D^x,D^n,t)~|~ t \in D^n, \mbox{ and } t \in \mc{CS}(D^n,D^x,\mc{Q})\}$
belongs to $\nit{PTIME}$.
\boxtheorem
\end{proposition}
\begin{example}\label{ex:al2} (ex. \ref{ex:rc2cp} cont.)  Consider the query $\mc{Q}\!: \  \exists xy (P(x)
\land Q(x,y)) \lor \exists xy(P(x) \land R(x,y))$, and assume that for $D$, \ $D^n=\{P(a), R(a,c)\}$ and $D^x=\{P(e),Q(a,b)\}$. It holds $\mf{S}(D)=\{ \{P(a),Q(a,b)\}, \{P(a),
R(a,c)\} \}$.  Since  $\{P(a)\} \subseteq \{P(a),
R(a,c)\}$, \ $\mf{S}^n(D)=\{\{P(a)\}\}$. So, $P(a)$ is the only actual cause for $ \mc{Q}$.
\boxtheorem
\end{example}
\vspace{-1mm}
\subsection{Contingency sets for unions of conjunctive queries}\label{sec:disjcausesCont} \vspace{-1mm}
 It is possible to develop a (naive) algorithm  that accepts as input an instance
$D=D^n \cup D^x$, and a UBCQ $\mc{Q}$,  and returns
$\nit{CS}(D, D^n,\mc{Q})$, and also, for
each $t \in \nit{CS}(D, D^n,\mc{Q})$, \ its (set of) S-minimal contingency sets $\mc{CT}(D,D^n,\mc{Q},t)$.
The basis for the algorithm is a correspondence between the actual causes for $\mc{Q}$ with their
contingency sets and a {\em hitting-set problem}.\footnote{If $\mc{C}$ is a collection of non-empty subsets of a set $S$, a subset $S' \subseteq S$ is a {\em hitting set} for
    $\mc{C}$  if, for every  $C \in \mc{C}$, $C \cap S' \neq \emptyset$. \ $S'$ is an S-minimal HS if no proper subset of
    it is also an HS. $S$ is a minimum HS is it has minimum cardinality.}

More precisely, for a fixed UBCQ $\mc{Q}$,  consider the {\em hitting-set framework} $\mf{H}^n(D) = \langle D^n, \mf{S}^n(D)\rangle$, with  $\mf{S}^n(D)$ as in Definition
\ref{def:hsStuff}. Different decision problems can be imposed on it.
The  S-minimal {\em hitting sets} (HSs) for $\mf{H}^n(D)$ correspond to actual causes with their S-minimal contingencies for $\mc{Q}$. Most responsible causes for $\mc{Q}$ are in correspondence with minimum hitting sets  for $\mf{H}^n(D)$.
 Notice that these hitting sets are all subsets of $D^n$. \vspace{-1mm}
\begin{proposition}\label{pro:UBCQCauses} \em
For $D=D^x \cup D^n$ and  a UBCQ $\mc{Q}$, it holds: \ (a) $t$ is an actual cause for $\mc{Q}$ with S-minimal contingency set $\Gamma$ iff  $\Gamma \cup \{t\}$ is an S-minimal HS for  $\mf{H}^n(D)$. \ (b) $t$ is a most responsible actual cause for $\mc{Q}$ with C-minimal contingency set $\Gamma$ iff  $\Gamma \cup \{t\}$ is a minimum HS  for $\mf{H}^n(D)$.\boxtheorem
\end{proposition}\vspace{-1.5mm}
\begin{example}\label{ex:al1} (ex. \ref{ex:rc2cp} and \ref{ex:al2} cont.) \  $D$ and $\mc{Q}$ are as before, but we now all tuples are endogenous.
 Here,
$\mf{S}(D)= \mf{S}^n(D) = \{ \{P(a),Q(a,b)\}, \{P(a),$
$R(a,c)\} \}$. $\mf{H}^n(D)$ has two S-minimal HSs:
$H_1=\{P(a)\}$ and $H_2=\{Q(a,b), R(a,c)\}$. Each of them implicitly contains an
actual cause (any of its elements)  with an S-minimal  contingency set (what's left after removing the actual cause). $H_1$ is also the C-minimal hitting set, and contains the most responsible actual cause,
$P(a)$.
\boxtheorem
\end{example}

\begin{rmk} \label{rem:hs}\em For $\mf{H}^n(D)=\langle D^n, \mf{S}^n(D)\rangle$, $\mf{S}^n(D)$ can be computed in PTIME, and its elements are bounded  in size by $|\mc{Q}|$,  which is the
maximum number of atoms in one of $\mc{Q}$'s disjuncts. This is a special kind of hitting-set problems. For example, deciding if there is a hitting set of size at most $k$ as been called the $d$-{\em hitting-set problem} \cite{NiedermeierR03}, and $d$ is the bound on the size of the sets in the set class. In our case, $d$ would be $|\mc{Q}|$. \boxtheorem
\end{rmk}

\vspace{-1mm}
\subsection{Causality, repairs and consistent answers} \label{sec:cqa}
\vspace{-1mm}
Corollary \ref{col:sr&cp} and  Proposition \ref{pro:UBCQCauses} can be used to compute repairs. If the classes of S- and C-minimal HSs for $\mf{H}^n(D)$ (with $D^n = D$) are available, computing S- and C-repairs will be in PTIME
in the sizes of those classes. However, it is well known that computing minimal HSs is a complex problem. Actually, as Example \ref{ex:repsVSconts} implicitly shows, we can have exponentially many of them in  $|D|$; so
 as exponentially many minimal repairs for a $D$  wrt. a denial constraint.\footnote{An example of this kind for FDs is given in \cite{Arenas03}. However, FDs form a special class of
 DCs that involve equality. Consequently, their violation views involve inequality.} So, the complexity of contingency sets computation is in line with the complexities of
 computing hitting sets and repairs.

The  computation of causes, contingency sets, and most responsible causes via minimal/minimum HS computation can then be used to compute repairs and decide about repair questions. Since the
 HS problems in our case are of the $d$-hitting set kind, good algorithms and approximations for the latter (cf. Section \ref{sec:fpt}) could be used in the context of repairs (all this via Corollary \ref{col:sr&cp} and Proposition \ref{pro:UBCQCauses}).

Consider an instance $D$ (with all tuples endogenous) and a set $\Sigma$ of DCs.
\ For the disjunctive violation view $V^\Sigma$, the following result is obtained from Propositions \ref{pro:sr&cp} and \ref{pro:cr&mrp}, and Corollary \ref{col:sr&cp}. \vspace{-1mm}
\begin{corollary}\label{col:consinf} \em
For an instance $D$ and set DCs
$\Sigma$, it holds: \ (a)  For every
$t \in \mc{CS}(D,V^\Sigma)$, there is an S-repair that does not contain $t$.
\ \ (b)  For every
$t \in \mc{MRC}(D, V^{\Sigma})$, there is a C-repair that does not contain $t$. \ \
(c) For every $D' \in \nit{Srep}(D,\Sigma)$ and $D'' \in \nit{Crep}(D,\Sigma)$, \
$D \smallsetminus D' \subseteq \mc{CS}(D,V^\Sigma)$ and $D \smallsetminus D'' \subseteq \mc{MRC}(D,V^\Sigma)$.
\boxtheorem
\end{corollary}
For a
projection-free, and a possibly non-boolean CQ $\mc{Q}$,
we are interested in its consistent answers from $D$ wrt. $\Sigma$. For example, for
$\mc{Q}(x,y,z)\!: \ R(x,y) \wedge S(y,z)$, the S-consistent (C-consistent) answers would be of the form
$(a,b,c)$, where $R(a,b)$ and $S(b,c)$ belong to
all S-repairs (C-repairs) of $D$. From Corollary \ref{col:consinf}, $(a,b,c)$ is an S-consistent, resp. C-consistent, answer iff $R(a,b)$ and $S(b,c)$ belong to $D$, but they are  not
actual causes, resp. most responsible actual causes, for  $ V^\Sigma$.

\ignore{A similar argument, appealing this time to \red{Proposition
\ref{pro:cr&mrp}}, can be used for consistent query answering wrt. the C-repairs and
most responsible causes for $V^{\Sigma}$.}


\ignore{
    \comlb{The proposition below may hold for projection-free,
possibly non-boolean, conjunctive queries, e.g. $\nit{ans}(x,y,z)
\leftarrow R(x,y), S(y,z)$. If this the case,
how does the prop. below change? *Some* atom is NOT in $D
\smallsetminus \mc{CS}(D, V^{\kappa})$; or *all*? Change mutatis
mutandi for C-repairs.}
}

\begin{proposition}\label{pro:cqa} \em
For an instance $D$, a set of DCs $\Sigma$, and a projection-free
CQ $\mc{Q}(\bar{x})\!: \  P_1(\bar{x}_1) \wedge \cdots
\wedge P_k(\bar{x}_k)$:\ \ (a) $\bar{c}$ is an S-consistent answer iff, for each $i$,  $P_i(\bar{c}_i) \in (D \smallsetminus \mc{CS}(D, V^{\Sigma}))$.\ \ (b)
$\bar{c}$ is a C-consistent answer iff, for each $i$,
$P_i(\bar{c}_i) \in (D
\smallsetminus \mc{MRC}(D, V^{\Sigma}))$.
\boxtheorem
\end{proposition}
\begin{example}\label{ex:cqa1} (ex. \ref{ex:rc2cp} cont.)  Consider
  $\mc{Q}(x)\!: \ P(x)$.   We had
$\mc{CS}(D, V^\Sigma)$ $=\{P(a),Q(a,b),$ $R(a,c) \}$,
$\mc{MRC}(D, V^\Sigma)= \{P(a)\}$. Then,  $a$ is both an S- and a C-consistent answer. \boxtheorem
\end{example}
Notice that Proposition \ref{pro:cqa} can easily be
extended to conjunction of ground atomic queries. Actually, from it we obtain the following result that
will be useful later on. \vspace{-1mm}
\begin{corollary}\label{cor:cqa&cox} \em
Given $D$, a set of DCs $\Sigma$, the ground atomic query $\mc{Q}\!\!:  P(c)$ is C-consistently true
 if  $P(c) \in D$ and it is not a most responsible cause for $V^\Sigma$.\boxtheorem
\end{corollary}
\begin{example}\label{ex:cqa2} For  $D=\{P(a,b),R(b,c),R(a,d)\}$ and the DC $\kappa\!: \ \leftarrow P(x, y),R(y, z)$: \
$\mc{CS}(D,$ $V^{\kappa})=\mc{MRC}(D, V^{\kappa})=\{P(a,b),R(b,c)\}$.
From Proposition \ref{pro:cqa}, the ground atomic query
$\mc{Q}\!\!: R(a,d)$ is
both S- and C-consistently true in $D$ wrt. $\kappa$,
because, $D \smallsetminus \mc{CS}(D, V^{\kappa}) = D \smallsetminus
\mc{MRC}(D, V^{\kappa})= \{R(a,d)\}$.
\boxtheorem
\end{example}
The CQs considered in  Proposition \ref{pro:cqa} and its Corollary \ref{cor:cqa&cox} are not
the particularly interesting, but will use those results to obtain relevant results for causality later on, e.g.
Theorem \ref{the:cqa&ca&cox}.

\section{Diagnosis: Query Answer Causality and Repairs} \label{sec:MBDtoRep}
\vspace{-1mm}
Let $D = D^n\cup D^x$ be an instance for schema $\mathcal{S}$, and
$\mc{Q}\!: \exists \bar{x}(P_1(\bar{x}_1) \wedge \cdots \wedge P_m(\bar{x}_m))$, a  BCQ.
Assume $\mc{Q}$ is, possibly  unexpectedly, true in  $D$. So, for the associated
DC $\kappa(\mc{Q})\!: \forall \bar{x} \neg (P_1(\bar{x}_1) \wedge \cdots \wedge P_m(\bar{x}_m))$,
$D \not \models \kappa(\mc{Q})$. $\mc{Q}$ is our {\em observation}, for which we want to find explanations, using a consistency-based diagnosis approach.

For each predicate $P \in \mc{P}$, we introduce predicate $\nit{Ab}_P$, with the same arity as $P$. A tuple
in its extension is  {\em abnormal} for $P$.
The ``system description", $\nit{SD}$, includes, among other elements,
the original database, expressed in logical terms, and the DC being true ``under normal conditions".
More precisely, we consider the following {\em diagnosis problem},  $\mathcal{M}=(\nit{SD},D^n, \mc{Q})$, associated to $\mc{Q}$. The FO system description, $\nit{SD}$,
contains the following elements:

\vspace{1mm}
\noindent  (a) $\nit{Th}(D)$, which is
Reiter's logical reconstruction of $D$ as a FO theory \cite{Reiter82} (cf. Example \ref{ex:mbdaex5}).

\vspace{1mm}
\noindent (b) Sentence $\kappa(\mc{Q}){\!^\nit{Ab}}$, which is $\kappa(\mc{Q})$ rewritten as follows:\vspace{-1mm}
\begin{eqnarray}
\kappa(\mc{Q}){\!^\nit{Ab}}\!:  \ \forall   \bar{x}\neg (P_1(\bar{x}_1)  \wedge \neg \nit{Ab}_{P_1}(\bar{x}_1)  \wedge \cdots \wedge
P_m(\bar{x}_m) \wedge \neg \nit{Ab}_{P_m}(\bar{x}_m) ). \label{eq:ext}
\end{eqnarray}
\phantom{poto}

\vspace{-6mm}
\noindent
This formula can be refined by applying the abnormality predicate, $\nit{Ab}$,
to endogenous tuples only. For this we need to use additional auxiliary predicates $\nit{End}_{P}$, with the same arity of $P \in \mc{S}$, which contain the endogenous
tuples in $P$'s extension (see Example \ref{ex:mbdaex5}).

\vspace{1mm}
\noindent (c) \ \ignore{The sentence $\neg \kappa(\mc{Q}) \longleftrightarrow \mc{Q}$, where $\mc{Q}$ is the initial boolean query. (d)} The inclusion dependencies: \
$\forall \bar{x}(\nit{Ab}_P(\bar{x}) \rightarrow P(\bar{x}))$,  \ $\forall \bar{x}(\nit{End}_P(\bar{x}) \rightarrow P(\bar{x}))$, and $\forall \bar{x}(\nit{Ab}_P(\bar{x}) \rightarrow \nit{End}_P(\bar{x}))$, for each $P \in \mc{P}$.

\vspace{1mm}\noindent
The last entry,  $\mc{Q}$, in $\mathcal{M}$  is the {\em observation}, which together with \nit{SD} will produce  and inconsistent theory,
because  we make the initial and explicit assumption that all the abnormality predicates are empty (equivalently, that  all tuples are normal), i.e. we
consider, for each predicate $P$, the sentence\footnote{Notice that these can also be seen as DCs, since they can be written as $\forall \bar{x} \neg \nit{Ab}_P(\bar{x})$.}\vspace{-2mm}
\begin{equation} \label{eq:default}
\forall \bar{x}(\nit{Ab}_P(\bar{x}) \rightarrow \mbox{\bf false}),
\end{equation}
\phantom{poto}

\vspace{-7mm}
\noindent where, {\bf false} is a propositional atom that is always false.
Actually, the second entry in $\mc{M}$
tells us how we can restore consistency, namely by (minimally) changing the abnormality condition on tuples in $D^n$. In other words, the rules (\ref{eq:default})
are subject to qualifications: some endogenous tuples may be abnormal. Each diagnosis  shows an S-minimal set of endogenous tuples that are abnormal.

\begin{example}\label{ex:mbdaex5} (ex. \ref{ex:cfex1} cont.) For the instance $D=\{S(a_3),$ $S(a_4),$ $R(a_4,a_3) \}$, with $D^n$ $=$ $\{S(a_4),$ $S(a_3)\}$, consider the diagnostic problem
 $\mathcal{M}=( \nit{SD},\{S(a_4),S(a_3)\},$ $ \mc{Q})$, with
$\nit{SD}$ containing:

\vspace{1mm}\noindent
(a) Predicate completion axioms: \ $\forall x y (R(x,y) \leftrightarrow x= a_4 \wedge y = a_3)$, \ $\forall x(S(x) \leftrightarrow x = a_3 \vee x = a_4)$,\\ \hspace*{5mm}$\forall x y (\nit{End}_R(x,y)$ $\leftrightarrow \mbox{\bf false})$,
 \ $\forall x(\nit{End}_S(x) \leftrightarrow x = a_3 \vee x = a_4)$.

Unique names assumption: $a_4 \neq a_3$.

\vspace{1mm} \noindent
(b) $\kappa(\mc{Q})^{\!\nit{Ab}}\!: \ \forall x y \neg ( S(x) \land \nit{End}_S(x) \land \neg   \nit{Ab}_S(x) \land  R(x, y) \land \nit{End}_R(x,y) \land \neg   \nit{Ab}_R(x, y) \ \land$\\
\hspace*{2.8cm}$S(y) \land \neg   \nit{Ab}_S(y))$.

\vspace{1mm} \noindent
(c) $\forall x y(\nit{Ab}_R(x,y) \rightarrow R(x,y))$, \ $\forall x(\nit{Ab}_S(x) \rightarrow S(x))$, $\forall x y(\nit{End}_R(x,y) \rightarrow R(x,y))$,\\
\hspace*{5mm}$\forall x(\nit{End}_S(x) \rightarrow S(x))$, $\forall x y(\nit{Ab}_R(x,y) \rightarrow \nit{End}_R(x,y))$, \ $\forall x(\nit{Ab}_S(x) \rightarrow \nit{End}_S(x))$.

\vspace{1mm}
The  normality assumptions for tuples are: \ $\forall x y(\nit{Ab}_R(x,y)$ $\rightarrow \mbox{\bf false})$, \ $\forall x(\nit{Ab}_S(x) \rightarrow \mbox{\bf false})$. \boxtheorem
\end{example}
\vspace{0.5mm}Now, the observation is $\mc{Q}$ (being true), obtained by evaluating query $\mc{Q}$ on (theory of) $D$. In this case, $D \not \models \kappa(\mc{Q})$. Since all the abnormality predicates are assumed to
be empty, $\kappa(\mc{Q})$ is equivalent to $\kappa(\mc{Q})^\nit{Ab}$, which also becomes false wrt $D$. As a consequence,  $\nit{SD} \cup \{(\ref{eq:default})\} \cup \{\mc{Q} \}$ is an inconsistent FO theory.
A
 diagnosis is a set of endogenous tuples that, by becoming abnormal, restore consistency.
 \begin{definition}\label{def:diag}
 (a) A {\em diagnosis} for $\mc{M}$ is a $\Delta \subseteq D^n$, such that
\ $\nit{SD} \cup \{\nit{Ab}_P(\bar{c})~|~P(\bar{c}) \in \Delta\} \cup \{\neg \nit{Ab}_P(\bar{c})~|~P(\bar{c}) \in D \smallsetminus \Delta\} \cup \{\mc{Q}\}$ \ is consistent.
\ (b)
$\mc{D}(\mc{M},t)$ denotes the set of S-minimal diagnoses for $\mc{M}$ that contain a tuple $t \in D^n$. \ (c) $\mc{MCD}(\mc{M},t)$ denotes the set of  C-minimal diagnoses
in $\mc{D}(\mc{M},t)$. \ignore{$\mc{M}$ that contain a tuple $t \in D^n$ and have the minimum cardinality (among those diagnoses that contain $t$)} \boxtheorem
\end{definition}
By definition,
$\mc{MCD}(\mc{M},t) \subseteq \mc{D}(\mc{M},t)$. Diagnoses for $\mc{M}$
and actual causes for $\mc{Q}$ are related.

\begin{proposition}\label{pro:ac&diag} \em
Consider  $D= D^n \cup D^x$, a BCQ $\mc{Q}$, and the diagnosis problem $\mc{M}$ associated to $\mc{Q}$. Tuple $t \in D^n$ is an actual cause for $\mc{Q}$
iff $\mc{D}(\mc{M},t) \not = \emptyset$. \boxtheorem
\end{proposition}
The responsibility of an actual cause $t$ is determined by the cardinality of the diagnoses in $\mc{MCD}(\mc{M},t)$.
\begin{proposition}\label{pro:r&diag} \em
For  $D= D^n \cup D^x$, a BCQ $\mc{Q}$, the associated diagnosis problem $\mc{M}$, and a tuple $t \in D^n$, it holds: \
(a) $\rho_{_{\!D\!}}(t)=0$ iff $\mc{MCD}(\mc{M},t) = \emptyset$.
\ (b) Otherwise, $\rho_{_{\!D\!}}(t)=\frac{1}{|s|}$, where $s \in \mc{MCD}(\mc{M},t)$. \boxtheorem
\end{proposition}
\begin{example}\label{ex:mbdaex6}   (ex. \ref{ex:mbdaex5} cont.)  $\mc{M}$ has two diagnosis:
$\Delta_1=\{ S(a_3)\}$ and $\Delta_4=\{ S(a_4)\}$.
\ Here, $\mc{D}(\mc{M},S(a_3))= \mc{MCD}(\mc{M}, S(a_3))=\{ \{ S(a_3)\}\}$ and $\mc{D}(\mc{M}, S(a_4))=\mc{MCD}(\mc{M}, S(a_4))=\{\{ $ $S(a_4)\}\}$. From Propositions
\ref{pro:ac&diag} and \ref{pro:r&diag}, $S(a_3)$ and $S(a_4)$ are actual cases, with responsibility $1$. \boxtheorem
\end{example}
In consistency-based diagnosis, minimal
diagnoses can be obtained as S-minimal HSs of the collection of S-minimal {\em conflict sets}  (cf. Section \ref{sec:prel}) \cite{Reiter87}.  In our case, conflict sets are S-minimal
sets of endogenous tuples  that, if not abnormal (only endogenous ones can be abnormal), and together, and possibly in combination with exogenous tuples, make (\ref{eq:ext}) false.
It is easy to verify that the conflict sets of $\mc{M}$ coincide with the sets in $\mf{S}(D^n)$ (cf. Definition \ref{def:hsStuff} and
Remark \ref{rem:hs}).  As a consequence, conflict sets for $\mc{M}$ can be computed in PTIME, the HSs for $\mc{M}$ contain actual causes for $\mc{Q}$, and the HS problem
for the diagnosis problems is of the $d$-hitting-set kind. The connection between consistency-based diagnosis and
causality allows us, in principle,  to apply techniques for the former, e.g. \cite{Feldman10, Mozetic94}, to the latter.

\begin{example}\label{ex:mbdGrep}  (ex. \ref{ex:mbdaex5} cont.)
The diagnosis problem  $\mathcal{M}=( \nit{SD},\{S(a_4),S(a_3)\},$ $ \mc{Q})$ gives rise to the HS framework $\mf{H}^n(D) = \langle  \{S(a_4),S(a_3)\}, \{\{ (S(a_3), S(a_4) \}\}\rangle$, with  $\{ S(a_3), S(a_4) \}$ corresponding to the
conflict set  $c=\{S(a_4),S(a_3)\}$. $\mf{H}^n(D)$ has two minimum HSs: $\{ S(a_3) \}$ and  $\{S(a_4) \}$, which are the S-minimal diagnosis for $\mathcal{M}$. Then, the two tuples are actual causes for $\mc{Q}$ (cf. Proposition\  \ref{pro:ac&diag}). From Proposition \ref{pro:r&diag}, $\rho_{_{\!D\!}}(S(a_3))= \rho_{_{\!D\!}}(S(a_4))= 1$.  \boxtheorem
\end{example}
The solutions to the diagnosis problem can be used for computing repairs.
\begin{proposition}\label{pro:diag} \em
Consider a database instance $D$ with only endogenous tuples, a set of DCs of the form $\kappa\!: \ \forall   \bar{x}\neg (P_1(\bar{x}_1)  \wedge  \cdots \wedge
P_m(\bar{x}_m)$, and their associated ``abnormality" integrity constraints\footnote{Notice that these are not denial constraints.}
in (\ref{eq:ext}) (in this case we do not need $\nit{End}_P$ atoms). Each S-minimal diagnosis $\Delta$ gives rise to an S-repair of $D$, namely $D_{\!\Delta} = D \smallsetminus \{P(\bar{c}) \in D~|~
\nit{Ab}_P(\bar{c}) \in \Delta\}$;  and every S-repair can be obtained in this way. Similarly, for C-repairs using C-minimal diagnoses. \boxtheorem
\end{proposition}
\begin{example}\label{ex:mbdaex6+} (ex. \ref{ex:mbdaex6} cont.)
The instance  $D=\{S(a_3),$ $S(a_4),$ $R(a_4,a_3) \}$ has three (both S- and C-) repairs  wrt. the DC $\kappa\!: \ \forall x y \neg ( S(x)  \land  R(x, y) \land  S(y))$, namely $D_1 = \{S(a_3) \}$,
$D_2 = \{S(a_4) \}$, and $D_3 = \{ R(a_4,a_3)\}$. They can be obtained as $D_{\!\Delta_1}, D_{\!\Delta_2},$ $ D_{\!\Delta_3}$ from the only (S- and C-) diagnoses,  $\Delta_1=\{ S(a_3)\}$, $\Delta_4=\{ S(a_4)\}$, $\Delta_3=\{R(a_4,a_3)\}$, resp.
\boxtheorem
\end{example}
The kind of diagnosis problem we introduced above can be formulated as a {\em preferred-repair problem} \cite[sec. 2.5]{2011Bertossi} (see \cite{chomicki12} for a general approach to
prioritized repairs). For this, it is good enough to
materialize tables for the auxiliary predicates $\nit{Ab}_P$ and $\nit{End}_P$, and consider the DCs of the form  (\ref{eq:ext}) (with the $\nit{End}_P$ atoms if not all
tuples are endogenous), plus the DCs (\ref{eq:default}). The initial extensions for the $\nit{Ab}_P$ predicates are empty. If $D$ is inconsistent wrt. this set of DCs, the S-repairs that are obtained by only {\em inserting} endogenous tuples into the
extensions of the $\nit{Ab}_P$ predicates correspond to S-minimal diagnosis, and each S-minimal diagnosis can be obtained in this way.

\vspace{-1mm}
\section{Complexity Results}\label{sec:MBDcomx}
\vspace{-1mm}
There are three main computational problems in database causality. For a BCQ $\mc{Q}$ and database $D$, they are: \
(a) The {\em causality problem} (CP) that is about computing the actual causes for $\mc{Q}$. \ (b) The
{\em responsibility problem} (RP) that is about computing the responsibility $\rho_{_{\!D\!}}(t)$ of a given actual cause $t$. Since a tuple that is not
an actual cause has responsibility $0$, the latter problem subsumes the former. \ (c) Computing the {\em most responsible actual causes} (MRC).
These problems have corresponding {\em decision versions}. Both CP and its decision version, CPD,
are solvable in polynomial time \cite{Meliou2010a}, which can be extended to UBCQs (cf. Proposition \ref{pro:UBCQCausesindirect}).
We consider the  decision version of the second  problem.
\begin{definition}  \label{def:resp}   For  a BCQ $\mc{Q}$,
the {\em responsibility decision problem} (RPD) is (deciding about membership of) \ \
 $\mathcal{RPD}(\mc{Q})=\{(D^x,D^n,t,v)~|~ t \in D^n, v \in \{0\} \cup \{\frac{1}{k}~|~k \in \mathbb{N}^+\},$  $D:=D^x \cup D^n \models \mc{Q}$ \ and \ $\rho_{_{\!D\!}}(t) > v  \}$. \boxtheorem
\end{definition}
 The complexity analysis of RPD in \cite{Meliou2010a} is restricted to conjunctive queries without self-joins, for which
a dichotomy result holds: \ depending on the syntactic structure of a query, RPD is either in PTIME or is NP-hard.  Here, we generalize the complexity analysis for RPD to general CQs.

We will also investigate the decision version, MRCD, of MRC, i.e. about deciding most responsible actual causes. This is a natural problem, because actual causes with the highest responsibility tend
to provide most interesting explanations for query answers
\cite{Meliou2010a,Meliou2010b}.
\begin{definition}  \label{def:mracp}   For a BCQ $\mc{Q}$, the {\em most responsible cause decision problem}
 is  $\mc{MRCD}(\mc{Q})$ $=\{(D^x,D^n,t)~|~ t \in D^n \mbox{ and } 0 < \rho_{_{\!D\!}}(t) \mbox{ is a maximum for } D: =D^x \cup D^n\}$.
\boxtheorem
\end{definition}
We start by analyzing a more basic decision problem:  {\em S-minimal contingency checking} (MCCD).
\begin{definition}  \label{def:cusp}  For a BCQ $\mc{Q}$,  $\mc{MCCD}(\mc{Q}):=\{(D^x,D^n,t,\Gamma)~|~\Gamma \in \mc{CT}(D^n \cup D^x,D^n,$ $\mc{Q},t)\}$ \ (cf.  (\ref{eq:ct})).
\boxtheorem
\end{definition}
Due to the results in Sections \ref{sec:causfrepair} and \ref{sec:repairfcauses}, it clear that there is a close connection between MCCD and  the {\em S-repair checking} problem in consistent query answering \cite[chap. 5]{2011Bertossi},
 about deciding if instance $D'$ is an S-repair of instance $D$ wrt. a set of integrity constraints. Actually,
the following result is obtained from the membership of the S-repair checking problem of LOGSPACE for DCs \cite[prop. 5]{Afrati09}.
\begin{proposition}\label{pro:CSPCcpx} \em For  a BCQ $\mc{Q}$,
$\mc{MCCD}(\mc{Q}) \in  \nit{PTIME}$. \boxtheorem
\end{proposition}
We could also consider the decision problem defined as in Definition \ref{def:cusp}, but with C-minimal $\Gamma$. We will not
use results about this problem in the following. Furthermore, its connection with the C-repair checking problem is less direct. As one can see from Section \ref{sec:causfrepair},
 C-minimal contingency sets correspond to a repair semantics somewhere between the S-minimal and C-minimal repair semantics (a subclass of \nit{Srep}, but a superclass  of \nit{Crep}): \ It is about an S-minimal repair with minimum cardinality that does not contain a particular tuple.

Now we establish that RPD is \nit{NP}-complete for CQs in general. The \nit{NP}-hardness is shown in \cite{Meliou2010a}. Membership of \nit{NP}
is obtained using Proposition \ref{pro:CSPCcpx}.
\begin{theorem}\label{the:RP(D)cmx} \em
(a) For every BCQ $\mc{Q}$,  $\mathcal{RPD}(\mc{Q}) \in  \nit{NP}$. (b) \cite{Meliou2010a} \ There are CQs $\mc{Q}$ for which $\mathcal{RPD}(\mc{Q})$  is  \nit{NP}-hard. \boxtheorem
\end{theorem}
In order to better understand the complexity of the problem, RP, of computing responsibility, we will investigate the {\em functional}, non-decision version of the problem.

The main source of complexity when computing responsibilities is related to the hitting-set problem  associated to $\mf{H}^n(D) = \langle D^n, \mf{S}^n(D)\rangle$ in Remark \ref{rem:hs}. In this case, it is about computing the cardinality of a minimum hitting set that contains a given vertex (tuple) $t$. That  this is
a kind of {\em $d$-hitting-set problem} \cite{NiedermeierR03} will be useful in Section \ref{sec:fpt}.

Our responsibility problem  can also be seen
as a {\em vertex cover problem} on the {\em hypergraph} $\mf{G}^n(D) = \langle D^n, \mf{E}^n(D)\rangle$ associated to $\mf{H}^n(D) = \langle D^n, \mf{S}^n(D)\rangle$. In it,
the set of hyperedges $\mf{E}^n(D)$ coincides with  the collection  $\mf{S}^n(D)$. Determining the responsibility of a tuple $t$ becomes the problem on hypergraphs of determining the size of a minimum
vertex cover (VC)\footnote{A set of vertices is a VC for a hypergraph if it intersects every hyperedge. Obviously, when we talk of {\em minimum} VC,
we are referring to minimal in cardinality.} that contains  vertex $t$ (among all VCs that contain the vertex).  Again, in this problem the hyperedges are
bounded  by $|\mc{Q}|$.\footnote{We  recall that
repairs of databases wrt. DCs can be characterized as maximal independent sets of {\em conflict hypergraphs} (conflict graphs in the case of FDs) whose vertices are the database tuples, and hyper-edges connect tuples
that together violate a DC \cite{Arenas03,Chomicki05}.}

\begin{example}\label{ex:hyperexm}  For $\mc{Q}\!: \exists xy(P(x)\wedge R(x,y)\wedge P(y))$, and $D= D^n=\{P(a), P(c), R(a,c),$  $R(a,a)\}$, \ $\mf{S}(D)=\mf{S}^n(D)=\{ \{P(a), R(a,a)\}, \{P(a), P(c), R(a,c)\} \}$.  $D$
is the set of vertices of hypergraph $\mf{G}^n(D)$, and its hyperedges are $\{P(a), R(a,a)\}$, $\{P(a), P(c), R(a,c)\}$. The following are the minimal VCs: $vc_1=\{ P(a)\}$, $vc_2=\{ P(c), $ $R(a,a)\}$, $vc_3=\{  R(a,a), R(a,c)\}$. Then, $ P(a)$ is an actual cause with responsibility $1$. The other tuples are actual causes with responsibility $\frac{1}{2}$.
\boxtheorem
\end{example}

\vspace{1mm}\noindent {\em To simplify the presentation, we will formulate and  address our  computational problems as problems for graphs (instead
of hypergraphs). However, our results still hold for hypergraphs \cite{icdt07}. Actually, the following {\em representation lemma} holds.}


\begin{lemma}\label{lemma:resclx} \em
There is a fixed database schema $\mathcal{S}$ and a BCQ $\mc{Q} \in L(\mathcal{S})$, without built-ins, such that, for every graph $G=(V,E)$ and  $v \in V$, there is
an instance $D$ for $\mathcal{S}$ and a tuple $t \in D$, such that the size of a minimum VC of $G$ containing
$v$ equals the responsibility of $t$ as an actual cause for $\mc{Q}$.\boxtheorem
\end{lemma}
Having represented our responsibility problem as a graph-theoretic problem, we first consider the following {\em membership minimal VC problem} (MMVC):  Given a graph $G=(V,E)$, a vertex $v \in V\!$, determine the size of a minimum VC of $G$ that contains $v$.

\begin{lemma}\label{lemma:MMVCand} \em
Given a graph $G$ and a vertex $v$ in it,  there is a graph $G'$ extending $G$ that can be constructed in polynomial time in $|G|$, such that the size of
a minimum VC for $G$ that contains $v$ and the size  of a minimum VC for $G'$ coincide.\boxtheorem
\end{lemma}
From this lemma and the $\nit{FP}^{\nit{NP(log} (n))}$-completeness of determining the size of a maximum clique in a graph  \cite{Krentel88}, we obtain:

\begin{proposition}\label{pro:MMVCcml} \em
MMVC problem for graphs is $\nit{FP}^{\nit{NP(log} (n))}$-complete.
\boxtheorem
\end{proposition}
From Lemma \ref{lemma:resclx} and Proposition \ref{pro:MMVCcml} we obtain the complexity result for RP. Membership can also be obtained from Theorem \ref{the:RP(D)cmx}.
\begin{theorem}\label{the:r&diag} \em (a)  For every BCQ without built-ins, $\mc{Q}$, computing the responsibility of a  tuple as a cause for $\mc{Q}$ is in $\nit{FP}^{\nit{NP(log} (n))}\!\!\!$. \ (b)
There is~a database schema and a BCQ $\mc{Q}$, without built-ins, such that computing the responsibility of a tuple as a cause for $\mc{Q}$ is $\nit{FP}^{\nit{NP(log} (n))}$-complete.
\boxtheorem
\end{theorem}
Now we address the most responsible causes problem, MRCD. We use the connection with consistent query answering of Section \ref{sec:cqa}, namely Corollary \ref{cor:cqa&cox}, and the
$P^{\nit NP(log(n))}$-completeness of consistent query answering under the C-repair semantics for  queries that are conjunctions of ground atoms and a particular DC  \cite[theo. 4]{icdt07}.
\begin{theorem}\label{the:cqa&ca&cox} \em (a)  For every BCQ without built-ins, $\mc{MRCD}(\mc{Q}) \in P^\nit{NP(log(n))}\!\!$. \ (b)
There is a database schema and a BCQ $\mc{Q}$, without built-ins, for which $\mc{MRCD}(\mc{Q})$ is $P^\nit{NP(log(n))}$-complete. \boxtheorem
\end{theorem}
From  Proposition \ref{pro:UBCQCauses} and the $\nit{FP}^\nit{NP(log(n))}$-completeness of determining the size of  C-repairs for DCs \cite[theo. 3]{icdt07}, we obtain the following for the computation of the highest responsibility value.

\ignore{
\combabak{We need the result from  \cite[theo. 3]{icdt07}, its not obtained directly from what we have so far. We need to change Lemma 2 to obtain it directly. }
\comlb{We should briefly explain why not.}
\combabak{ I am not sure what would you expect me to say here. How can I explain that I did not want to redo thing that already has been done.}
\comlb{NEW: I am not asking to explain that, but why it does not follow from that Lemma.}
\combabak{NEW: We may need to talk about this. I really don't know how to explain}       }

\begin{proposition}\label{pro:crepair&res&cox} \em (a)  For every BCQ without built-ins, computing the responsibility of the most responsible causes   is in $\nit{FP}^\nit{NP(log(n))}$. \ (b)
There is a database schema and a BCQ $\mc{Q}$, without built-ins, for which computing the responsibility of the most responsible causes is  $\nit{FP}^\nit{NP(log(n))}$-complete. \boxtheorem
\end{proposition}

\subsection{FPT of responsibility}\label{sec:fpt}\vspace{-1mm}
 We need to cope with the intractability of computing most responsible causes.
The area of {\em fixed parameter tractability} (FPT) \cite{flum} provides tools to attack this problem. In this regard, we recall that a decision problem with inputs
of the form $(I, p)$, where $p$ is a distinguished
parameter of the input, is fixed parameter tractable (or belongs to
the class FPT), if it can be solved in time $O(f(|p|) \cdot |I|^c)$, where $c$ and the
hidden constant do not depend on $|p|$ or $|I|$, and $f$ does not depend on $|I|$.

In our case, the {\em parameterized version of the decision problem} $\mathcal{RPD}(\mc{Q})$ (cf. Definition \ref{def:resp}) is denoted with $\mathcal{RPD}^p(\mc{Q})$, and the distinguished parameter is $k$, such
that
$v = \frac{1}{k}$. That $\mathcal{RPD}^p(\mc{Q})$ belongs to FPT can be obtained from its formulation as a $d$-hitting-set problem ($d$ being the fixed upper bound on the size of the
sets in the set class); in this case about
deciding if there is a HS that contains the given tuple $t$ that has cardinality smaller that $k$. This problem  belongs to FPT.
\begin{theorem}\label{the:FPT} \em
For every BCQ $\mc{Q}$, $\mathcal{RPD}^p(\mc{Q})$ belongs to FPT, where the parameter is the inverse of the responsibility bound.
\boxtheorem
\end{theorem}
The proof of this result is interesting {\em per se}, and we sketch it here. First, there is a  PTIME
parameterized algorithm for the $d$-hitting-set problem about deciding if there is a HS of size at most  $k$ that runs in time
$O(e^k + n)$, with $n$ the size of the underlying set and  $e=d-1+o(d^{-1})$  \cite{NiedermeierR03}. In our case,  $n=|D|$, and $d = |\mc{Q}|$ \ (cf. also \cite{Fernau10}).

Now, to decide if the responsibility of a given tuple $t$ is greater than $v=\frac{1}{k}$, we consider the associated hypergraph $\mf{G}^n(D)$, and we decide
if it has a VC that contains $t$ and whose size is less than $k$. In order to answer this, we use Lemma \ref{lemma:MMVCand}, and  build the extended hypergraph $\mf{G}'$. The size of a minimum VC for $\mf{G}'$
gives the size of the
minimum VC of $\mf{G}^n(D)$ that contains $t$. If $\mf{G}^n(D)$ has a  VC that contains $t$ of size less than $k$, then $\mf{G}'$ has a VC of size less than $k$. If  $\mf{G}'$ has a VC of size less than $k$,  its minimum size for a VC is less than $k$. Since this minimum is the same as the size of a minimum VC for $\mf{G}^n(D)$ that contains $t$, $\mf{G}^n(D)$ has a VC of size less than $k$ that contains $t$. As a consequence,
it is good enough to decide if $\mf{G}'$ has a VC of size less than $k$. For this, we use the HS formulation of this hypergraph problem, and the already mentioned FPT algorithm.

This result and the corresponding algorithm show  that the higher the required responsibility degree, the lower the  computational effort needed to compute the actual causes with at least that level of responsibility. In other
terms, parameterized algorithms are effective for computing actual causes with high responsibility or most responsible causes. In general, parameterized algorithms are
very effective when the parameter is relatively small \cite{flum}.

Now, in order to compute most responsible causes, we could apply, for each actual cause $t$, the just presented FPT algorithm on the hypergraph $\mf{G}^n(D)$, starting with $k=1$, i.e. asking if there is
  VC of size less than $1$ that contains $t$. If the algorithm returns a positive result, then $t$  is a counterfactual cause, and has responsibility $1$. Otherwise, the algorithm will be launched with $k=2 , 3, \ldots, |D^n|$, until a positive result is returned. (The procedure can be improved through binary search on  $k=1, 2 , 3, \ldots,  m$, with $m$ possibly much smaller than $|D|$.)

The complexity results and algorithms provided in this section can be extend to UBCQs. This is due to Remark \ref{rem:ucq} and the construction of  $\mf{S}^n(D)$, which the results in this section build upon.  

For the $d$-hitting-set problem there are also efficient parameterized approximation algorithms \cite{Brankovic12}. They could be used to approximate the responsibility problem.
Furthermore,  approximation algorithms developed for the minimum VC
problem on bounded hypergraphs \cite{Halperin00,Okun05} should be applicable to approximate most responsible causes for query answers. Via the causality/repair connection (cf. Section \ref{sec:cqa}), it should be possible to
develop approximation algorithms to compute  S-repairs of particular sizes, C-repairs, and consistent query answers wrt. DCs.


\vspace{-1mm}
\subsection{The causality dichotomy's reflection on repairs}\label{sec:dichReps}
\vspace{-1mm}
In \cite{Meliou2010a} the class of {\em linear} CQs is introduced. For them, computing tuple responsibilities is tractable. Roughly speaking, a BCQ
is linear if its atoms can be ordered in a way
that every variable appears in a continuous sequence of atoms, e.g. $\mc{Q}_1\!:  \exists xvyu(A(x) \wedge S_1(x, v) \wedge S_2(v, y) \wedge R(y, u) \wedge S_3(y, z))$ is linear, but not $\mc{Q}_2\!:  \exists x y z(A(x) \wedge B(y) \wedge C(z) \wedge W(x, y, z))$, for which RPD is \nit{NP}-hard \cite{Meliou2010a}.  The class of BCQs for which computing responsibility (more precisely, our $\mc{RPD}$ decision problem) is tractable can be extended  to {\em weakly linear}.\footnote{Computing  sizes of minimum contingency sets is reduced to the max-flow/min-cut problem in a network.}
Now, the dichotomy result in \cite{Meliou2010a} says that for a BCQ $\mc{Q}$ without self-joins, RDP is tractable when $\mc{Q}$ is weakly-linear, but \nit{NP}-hard, otherwise.
Due to the causality/repair connection  of Section \ref{sec:repairfcauses}, we can obtain the following results for database repairs.

\vspace{-0.5mm}
\begin{theorem} \label{theo:dichotomy}\em
 (a) For single weakly-linear DCs, C-repair checking and deciding if the size of a C-repair is larger than a bound are both tractable.\footnote{A DC $\kappa$ is weakly-linear if the corresponding BCQ $V^\kappa$ is weakly-linear. In this way any adjective that applies to
 BCQs can be applied to DCs.}\\
 (b) For single, self-join free DCs $\kappa$, and the problem $\nit{RepSize}(\kappa)$ of deciding if there is a repair $D'$ for a given input instance $D$ and a tuple $t\in D$ with $|D'| \geq m$ and $t \not \in D'$,\footnote{More precisely,
  $D'$ is a subset of $D$ that satisfies $\kappa$. Here, $0 \leq m \leq n = |D|$.} the following dichotomy holds: (b1) If $\kappa$ is weakly-linear, $\nit{RepSize}(\kappa)$ is tractable. (b2) Otherwise, it is \nit{NP}-complete.
  \boxtheorem
\end{theorem}
This dichotomy result for repairs shows that interesting results
in one of the areas (causality, in this case) have counterparts in some of the others.
The form the reincarnation of the known result takes in the new
area (repairs, in this case) is interesting {\em per se}.

Notice that both problems in (a) in Theorem \ref{theo:dichotomy} may be intractable even for single DCs \cite{icdt07}. More specifically, C-repair checking can be \nit{coNP}-hard for single DCs  \cite{icdt07,Afrati09}. Actually, the single DC used in  \cite[lemma 4]{icdt07} is of the form
$\kappa\!: \ \leftarrow V(x), V(y), E(x,y,z)$, whose associated BCQ is not weakly-linear. As a matter of fact, this BCQ  is a \nit{NP}-hard for RDP \cite{Meliou2010a}.

\vspace{-1mm}
\section{Discussion and Conclusions}\label{sec:disc}
\vspace{-1mm}
In this research we have unveiled and formalized some first interesting  relationships between causality in databases, database repairs, and consistency-based diagnosis. These connections allow us to apply results and techniques developed for
each of them to the others. This is  particularly beneficial for causality in databases, where still a  limited number of results and techniques have been obtained or developed.

The connections we established here inspired complexity results for causality, e.g. Theorems \ref{the:r&diag} and \ref{the:cqa&ca&cox}, and were used to prove them.
We appealed to several non-trivial results (and the proofs thereof) about repairs/CQA obtained in \cite{icdt07}.
It is also the case that the
well-established hitting-set approach to diagnosis inspired
a similar  approach to causal responsibility, which in its turn allowed
us to obtain results about its fixed-parameter tractability.   It is also the case that diagnostic reasoning, as a form of non-monotonic reasoning, can provide a  solid foundation  for causality in databases and query answer explanation, in general \cite{Cheney09b,Cheney11}.

Our work creates a theoretical basis for deeper and
mathematically more complex investigations. In particular,
it also opens interesting research directions, some of which are briefly discussed below.

\vspace{1mm}
   \noindent{\bf Preferred causes for queries.} \
   In Section \ref{sec:causfrepair} we characterized causes and most responsible causes in terms of S-repairs and C-repairs, resp. This could be generalized
   by using the notion of {\em preferred repair} \cite{chomicki12}. These are repairs whose minimization correspond to a {\em priority relationship}, $\preceq$,  between instances.
   Let assume it defines a corresponding class of preferred repairs, $\preceq\!\!\nit{Rep}$.
Inspired by (\ref{eq:df}), we can define, for a BCQ $\mc{Q}$: \ $\mc{DF}^\preceq(D, D^n,\kappa(\mc{Q}), t):=\{ D \smallsetminus D'~|~ D' \in$ $\preceq\!\!\nit{Rep}(D,\kappa(\mc{Q})),~ t \in (D\smallsetminus D') \subseteq D^n\}$,
and, $t \in D^n$ is a \ $\preceq$-cause iff
$\mc{DF}^\preceq(D, D^n,\kappa(\mc{Q}), t) \not = \emptyset$. In this way, a whole class of preferences on causes can be introduced, which is natural problem \cite{Meliou2010b}.\footnote{In \cite{Meliou2011} the possibility of introducing weights
in the  partition is considered, in this way imposing  a form of preference on causes.}

\vspace{1mm}\noindent {\bf Endogenous repairs.} \ The partition of a database into endogenous and exogenous tuples may also be of interest in the context of repairs. Considering that we should
have more control on endogenous tuples than on exogenous ones, which may come from external sources, it makes sense to consider
{\em endogenous repairs}. They are obtained by updates (of any kind) on endogenous tuples. For example, in the case of DCs,
   endogenous repairs would be obtained by deleting endogenous tuples only. \ignore{For illustration, in Example \ref{ex:mbdaex6}, if $D^n$ $=$ $\{S(a_4),$ $S(a_3)\}$, then the
   only endogenous repairs would be $D_1$ and $D_2$.}
   If there are no repairs based on endogenous tuples, a preference condition could be imposed on repairs \cite{ihab12,chomicki12}, privileging those that change exogenous the least. (Of course,
   it could also be the other way around, that is we may feel more inclined to change exogenous tuples than our endogenous ones.)

  As a further extension, it could be possible to assume that combinations of (only) exogenous tuples never violate the integrity constraints, which could be checked
   at upload time. In this sense, there would be a part of the database that is considered to be consistent, while the other is subject to possible repairs. (For slightly
   related research, see  \cite{greco14}.)

 \ignore{ Actually, going a bit further, we could even consider the relations  in the database with an extra, binary  attribute, $N$, that is used to annotate if a tuple is
  endogenous or exogenous (it could be both), e.g. a tuple like $R(a,b, \nit{yes})$. integrity constraints could be annotated too, e.g. the ``exogenous" version of DC $\kappa$, could be
   $\kappa^E\!: \ \leftarrow P(x, y,\nit{yes}),R(y, z,\nit{yes})$, and could be assumed to be satisfied.   }

\vspace{1mm}\noindent{\bf Objections to causality.} \
Causality as introduced by Halpern and Pearl in \cite{Halpern01,Halpern05}, aka. HP-causality, is the basis for the notion of causality in \cite{Meliou2010a}. HP-causality has been the object of some criticism \cite{halpern14}, which is justified in some (more complex, non-relational) settings, specially due to the presence of different kinds of {\em logical variables} (or lack thereof). In our context
 the objections do not apply:  variables just  say  that a certain tuple belongs to the instance (or not); and for relational databases the closed-world assumption applies.
In \cite{halpern14}, the definition of HP-causality is slightly modified. In our setting, this modified definition does not change actual causes or their properties.

\ignore{
\noindent{\bf Open queries.} \ We have limited our discussion to boolean queries. It is possible to extend our work
to consider conjunctive queries with free variables, e.g. $\mc{Q}(x)\!: \exists yz(R(x,y) \wedge S(y,z))$. In this case,
a query answer would be of the form $\langle a\rangle$, for $a$ a constant, and causes would be found for such an answer.
In this case, the associated DC would be of the form $\kappa^{\langle a\rangle}\!: \ \leftarrow R(a,y), S(y,z)$, and
the rest would be basically as above.  }

\ignore{
\paragraph{\bf Algorithms and complexity.} \ Given the connection between causes and different kinds of repairs, we might take advantage
for causality of algorithms and complexity results obtained for database repairs. This is matter of our ongoing research. In this work, apart
from providing a naive algorithm for computing repairs from causes, we have not gone into detailed algorithm or complexity issues. The results
we already have in this direction will be left for an extended version of this work.
}

\vspace{1mm}
\noindent{\bf ASP specification of causes.} \  S-repairs can be specified by means of
   {\em answer set programs} (ASPs) \cite{tplp03,barcelo03}, and C-repairs too, with the use of weak program constraints \cite{tplp03}. This should allow for the
   introduction of ASPs in the context of causality, for specification and reasoning.
    There are also ASP-based specifications of diagnosis \cite{eiter99} that could be brought into a more complete picture.

\vspace{1mm}
\noindent{\bf Causes and functional dependencies, and beyond.} \ Functional dependencies are DCs with conjunctive violation views with
    inequality, and are still monotonic. There is much research on repairs and consistent query answering for functional dependencies, and more complex integrity constraints \cite{2011Bertossi}. In causality, mostly
    CQs without built-ins have been considered. The repair connection could be exploited  to obtain results for causality and CQs with inequality, and also other classes of queries.
    \ignore{It is possible that causality can be extended to conjunctive queries with built-ins through the repair connection; and also to non-conjunctive queries via repairs wrt.\ more complex
    integrity constraints.}

 \vspace{1mm}   \noindent {\bf View updates and abduction.} \ Abduction \cite{Console91,EiterGL97} is another form of model-based diagnosis, and is related to the subjects investigated in this work.The {\em view update problem}, about updating a database through views,
is a classical problem in databases that has been treated through abduction
\cite{Kakas90,Console95}. User knowledge imposed through view updates creates or reflects {\em uncertainty} about the base data, because alternative base instances may give an account
of the intended view updates.
The view update problem, specially in its particular form of of {\em deletion propagation}, has been recently related in \cite{benny12a,benny12b} to causality as introduced in
\cite{Meliou2010a}. (Notice only tuple deletions are used with violation views and repairs associated to DCs.)

 Database repairs are also related to the view update problem.
Actually, {\em answer set programs} (ASP) for database repairs \cite{barcelo03} implicity repair the database by updating intentional,
annotated predicates.
Even more, in \cite{lechen}, in order to protect sensitive information, databases are explicitly and virtually ``repaired" through secrecy views that specify the
information that has to be kept secret. These are prioritized repairs that have been specified via ASPs. Abduction has been explicitly applied to database repairs \cite{arieli}.
The deep interrelations between causality, abductive reasoning, view updates and repairs are  the objects of our ongoing research efforts \cite{buda14}.

\vspace{1mm}
\noindent {\bf Acknowledgments:} \ Research funded by NSERC Discovery, and
the NSERC Strategic Network on Business Intelligence (BIN).  Conversations  with Alexandra Meliou during Leo Bertossi's visit to U. of Washington in 2011 are much appreciated.
He is also grateful to Dan Suciu and  Wolfgang Gatterbauer for their hospitality. L. Bertossi is grateful to Benny Kimelfeld for stimulating conversations.
Part of the research was developed by L. Bertossi at {\em LogicBlox} and {\em The Center for Semantic Web Research} (Chile). Their support is much appreciated.

\appendix

\section{Appendix: \ Proofs of Results}\label{ap:proofs}

     \defproof{Proposition \ref{pro:UBCQCausesindirect}}{Assume $\mf{S}(D) =\{s_1, \ldots, s_m\}$, and there exists a $s \in\mf{S}^n(D)$ s.t. $t \in s$. Consider a set $\Gamma \subseteq D^n$ such that, for all $s_i \in\mf{S}^n(D)$ where $s_i \not = s$,  $\Gamma \cap s_i \not = \emptyset$ and  $\Gamma \cap s =\emptyset$. With such a $\Gamma$,
     $t$ is an actual cause for $\mc{Q}$ with contingency set $\Gamma$. So, it is good enough to prove that such $\Gamma$ always exists. In fact, since all subsets of $\mf{S}^n(D)$ are S-minimal, then, for each $s_i \in\mf{S}^n(D)$ with $s_i \not = s$, $s_i \cap s = \emptyset$. Therefore, $\Gamma$ can be obtained from the set of difference between each $s_i$ and $s$.

     Now, if $t$ is an actual cause for $\mc{Q}$, then there exist an S-minimal $\Gamma \in D^n$, such that  $D   \smallsetminus (\Gamma \cup\{t\}) \not \models \mc{Q}$, but  $D  \smallsetminus \Gamma \models \mc{Q}$. This implies that there exists an S-minimal subset of $s \in D$, such that $t \in s
 $ and $s \models \mc{Q}$. Due to the S-minimality of $\Gamma$, it is easy to see that $t$ is included in a subset of $\mf{S}^n(D)$. }

 \defproof{Proposition \ref{pro:UBCQCauses}}{Similar to the proof of Proposition \ref{pro:UBCQCausesindirect}.}

 \defproof{Propositions \ref{pro:ac&diag} and \ref{pro:r&diag}}{ It is easy to verify that the conflict sets of $\mc{M}$ coincide with the sets in $\mf{S}(D^n)$ (cf. Definition \ref{def:hsStuff}). The results obtained from the characterization of minimal diagnosis as minimal hitting sets of sets of conflict sets (cf. Section \ref{sec:prel} and \cite{Reiter87}) and Proposition \ref{pro:UBCQCauses}.}

 \defproof{Proposition \ref{pro:CSPCcpx}}{We provide
a  PTIME algorithm to decide if $(D^x,D^n,t,\Gamma) \in  \mc{MCCD}(\mc{Q})$. Consider $D$ and the DC $\kappa(\mc{Q})$  associated to $\mc{Q}$ (cf. Section \ref{sec:causfrepair}). $(D^x,D^n,t,\Gamma) \in  \mc{MCCD}(\mc{Q})$ iff $D \smallsetminus  (\{t\} \cup \Gamma)$ is an S-repair for $D$ (which follows from the proof of Proposition \ref{pro:c&r}).  Repair checking can be done in LOGESPACE \cite[prop. 5]{Afrati09}, therefore the decision can be made in PTIME.}

\defproof{Theorem \ref{the:RP(D)cmx}}{We describe
a non-deterministic PTIME algorithm to decide  RPD. Non-deterministically guess a subset $\Gamma \subseteq D^n$, return {\em yes} if $|\Gamma| < \frac{1}{v}$ and $(D^x, D^n, t, \Gamma ) \in  \mc{MCCD}$; otherwise return {\em no}. According to Proposition \ref{pro:CSPCcpx} this can be done in PTIME in data complexity.}

\defproof{Lemma \ref{lemma:resclx}}{ Consider a graph $G = (V, E)$, and assume the vertices
of $G$ are uniquely labeled. Consider the database schema with relations,
$\nit{Ver}(v_0)$ and $\nit{Edges}(v_1, v_2, e)$, and the conjunctive query $\mc{Q}\!: \exists v_1v_2e(\nit{Ver}(v_1) \wedge \nit{Ver} (v_2) \wedge \nit{Edges}(v_1, v_2, e))$.
$\nit{Ver}$ stores the vertices of $G$, and $\nit{Edges}$, the labeled edges. For each edge $(v_1, v_2) \in G$, $\nit{Edges}$ contains $n$ tuples of the form $(v_1, v_2, i)$, where $n$ is the
number of vertices in $G$. All the values in the third attribute of $\nit{Edges}$ are different, say from 1 to $n|E|$. The size of the database instance obtained through this
padding of $G$ is still polynomial in size.  It is clear that $D \models \mc{Q}$.

Assume $\nit{VC}$ is the minimum vertex cover of $G$ that contains the vertex $v$.  Consider the set of tuples $s= \{ \nit{Ver}(x)~|~x \in \nit{VC} \}$. Since  $v \in \nit{VC}$,    $s=s' \cup \nit{\{Ver}(v)\}$. Then, $D \smallsetminus (s' \cup \nit{Ver}(v)) \not \models Q$. This is because for every tuple $\nit{Edge}(v_i, v_j,k)$ in the instance, either $v_i$ or $v_j$ belongs to $VC$.\ignore{therefore, wither $\nit{Ver(v_i)}$ or $\nit{Ver(v_j)}$ belong to $s$}. Due to the minimality of $\nit{VC}$, $D  \smallsetminus s' \models \mc{Q}$.

Therefore,  tuple $\nit{Ver(v)}$ is an actual cause for $\mc{Q}$. Suppose, $\Gamma$ is  a C-minimal contingency set associated to  $\nit{Ver}(v)$. Due to the C-minimality  of
$\Gamma$, it entirely consists of tuples in $\nit{Ver}$. It holds that $D \smallsetminus (\Gamma \cup \{\nit{Ver}(v')\}) \not \models \mc{Q}$ and $D \smallsetminus \Gamma \models \mc{Q}$. Consider the set  $\nit{VC'}=\{x| \nit{Ver}(x) \in \Gamma\} \cup \{v'\}$. Since $D  \smallsetminus (\Gamma \cup \{\nit{Ver}(v')\}) \not \models \mc{Q}$, for every tuple $\nit{Edge}(v_i, v_j,k)$ in $D$, either $v_i \in VC' $ or  $v_j \in  VC'$. Therefore, $\nit{VC}'$ is a minimum vertex cover of $G$ that contains $v$. It holds that $\rho_{_{\!D\!}}(\nit{Ver}(v))=\frac{1}{1+|\Gamma|}$. So the size of a minimum vertex cover of $G$ that contains $v$ can be obtained from $\rho_{_{\!D\!}}(\nit{Ver}(v))$.}

  \defproof{Lemma \ref{lemma:MMVCand}}{The size of  $\nit{VC_G(v)}$, the minimum vertex cover of $G$ that contains the vertex $v$, can be computed from the size of $I_G$, the maximum independent set of $G$, that does not contains $v$. In fact,
  \begin{equation}\label{one}
  |\nit{VC_G(v)}|=|G|-|I_G|.
  \end{equation}
Since $I$ is a maximum independent set that does not contain $v$, it must contain
one of the adjacent vertices of $v$ (otherwise, $I$ is not maximum, and $v$ can be added to $I$). Therefore, $|\nit{VC_G(v)}|$ can be computed from the size of a maximum independent set $I$ that contains $v'$, one of the adjacent vertices of $v$.

Given a graph $G$ and a vertex $v'$ in it, a graph $G'$ that extends $G$ can be constructed in polynomial time in the size of $G$, such that there is a maximum independent set $I$ of $G$ containing $v'$ iff $v'$ belongs to
every maximum independent set of $G'$ iff the sizes of maximum independent sets for $G$ and $G'$ differ by one \cite[lemma 1)]{icdt07}.  Actually, the graph $G'$ in this lemma can be obtained by adding a new vertex $v''$ that is connected only to the neighbors of $v'$. Its holds:
\begin{eqnarray}
|I_G|&=&|I_G'|-1, \label{two}\\
|I_G'|&=&|G'|-|\nit{VC}_{G'}|, \label{three}
\end{eqnarray}
where $\nit{VC}_{G'}$ is a minimum vertex cover of $G'$.
From (\ref{one}), (\ref{two}) and (\ref{three}), we obtain:  $|\nit{VC_G(v)}|= |\nit{VC}_{G'}|$.}

 \defproof{Proposition \ref{pro:MMVCcml}}{We prove membership by describing
an algorithm in $\nit{FP}^{\nit{NP(log} (n))}$ for computing the size of the minimum vertex cover of a graph $G=(V,E)$ that contains a vertex $v \in V\!$.  We use  Lemma \ref{lemma:MMVCand}, and build the extended graph $G'$. The size of a minimum VC for $G'$ gives the size of the minimum VC of $G$ that contains $v$.  Since computing the maximum cardinality of a clique can be done in time $\nit{FP}^{\nit{NP(log} (n))}$  \cite{Krentel88}, computing a minimum vertex cover  can be done in the same time (just consider the complement graph).  Therefore, MMVC belong to $FP^{NP(log (n))}$.

Hardness can be obtained by a reduction from computing minimum vertex covers in graphs to MMVC. Given a graph $G$ construct the graph $G'$ as follows:  Add a vertex $v$ to $G$ and connect it to all vertices of $G$. It is easy to see that $v$ belongs to all minimum vertex covers of
$G'$. Furthermore,  the sizes of minimum vertex covers for $G$ and $G'$ differ by one. Consequently, the size of a minimum vertex cover of $G$
 can be obtained from the size of a minimum vertex cover of $G'$ that contains $v$. Computing the minimum vertex cover is $\nit{FP}^{\nit{NP(log} (n))}$-complete. This follows from the $\nit{FP}^{\nit{NP(log} (n))}$-completeness of computing the maximum cardinality of a clique in a graph  \cite{Krentel88}.}

\defproof{Theorem \ref{the:cqa&ca&cox}}{(a) We provide
an algorithm in $\nit{P}^{\nit{NP(log} (n))}$ to decide whether $(D^x,D^n,t) \in \mc{MRCD}(\mc{Q})$. Construct the hitting set framework $\mf{H}^n(D) = \langle D^n, \mf{S}^n(D)\rangle$ (cf. Definition \ref{def:hsStuff} and Remark \ref{rem:hs}) and its associated hypergraph $\mf{G}^n(D) = \langle D^n, \mf{E}^n(D)\rangle$, where,  $\mf{E}^n(D)$ coincides with the collection  $\mf{S}^n(D)$. It holds that $t$ is a most responsible cause for $\mc{Q}$ iff   $\mf{H}^n(D)$ has a C-minimal hitting set that contains $t$ (cf. Proposition \ref{pro:UBCQCauses}). Therefore, $t$ is a most responsible cause for $\mc{Q}$ iff $t$ belongs to some minimum vertex cover of $\mf{G}^n(D)$. Its is easy to see that $\mf{G}^n(D)$ has a minimum vertex cover that contains $t$  iff $\mf{G}^n(D)$ has a maximum independent set that does not contains $t$. Checking if $t$ belongs to all maximum independent set of $\mf{G}^n(D)$ can be done in  $P^\nit{NP(log(n))}$ \cite[lemma 2]{icdt07}. If $t$ belongs to all independent sets of $\mf{G}^n(D)$, then $(D^x,D^n,t) \not \in \mc{MRCD}(\mc{Q})$; otherwise $(D^x,D^n,t) \in \mc{MRCD}(\mc{Q})$. As a consequence, the decision can be made in time $P^\nit{NP(log(n))}$.

\vspace{1mm}
\noindent (b) The proof is by a reduction, via Corollary \ref{cor:cqa&cox}, from consistent query answering under the C-repair semantics for queries that are conjunctions of ground atoms, which was proved to be $P^\nit{NP(log(n))}$-complete in  \cite[theo. 4]{icdt07}. Actually, that proof (of hardness) uses a particular database schema $\mc{S}$ and a DC $\kappa$.  In our case, we can use the same schema $\mc{S}$ and the violation query $V^\kappa$ associated to $\kappa$ (cf. Section  \ref{sec:repairfcauses}).}

\defproof{Proposition \ref{pro:crepair&res&cox}}{(a) We describe
an algorithm in $\nit{FP}^{\nit{NP(log} (n))}$ that, given an instance $D=D^n \cup D^x$ and a BCQ $\mc{Q}$,  computes the responsibility of  most responsible causes for $\mc{Q}$. Consider the hypergraph $\mf{G}^n(D)$ as obtained in Theorem \ref{the:cqa&ca&cox}. The responsibility of most responsible causes for $\mc{Q}$ can be obtained from the size of the minimum vertex cover of $\mf{G}^n(D)$ (cf. Proposition \ref{pro:UBCQCauses}).  The size of the minimum vertex cover in a graph can be computed in $\nit{FP}^{\nit{NP(log} (n))}$, which is obtained from the membership of $\nit{FP}^{\nit{NP(log} (n))}$ of computing  the maximum cardinality of a clique  in graph \cite{Krentel88}. It is easy to verify that minimum vertex covers in hyprgraphs can be computed in the same time.

\vspace{1mm}
\noindent (b) This is by a reduction from the problem of determining the size of C-repairs for DCs  shown to be  $\nit{FP}^\nit{NP(log(n))}$-complete in \cite[theo. 3]{icdt07}. Actually, that proof (of hardness) uses a particular database schema $\mc{S}$ and a DC $\kappa$. In our case, we may consider the same schema $\mc{S}$ and the violation query
$V^{\kappa}$ associated to $\kappa$ (cf. Section  \ref{sec:repairfcauses}). The size of C-repairs for an inconsistent instance $D$ of the schema $\mc{S}$ wrt. $\kappa$ can be obtained from the responsibility of  most responsible causes for $V^{\kappa}$ (cf.  Corollary  \ref{col:sr&cp}).}

\defproof{Theorem \ref{theo:dichotomy}}{(a) We use Proposition \ref{pro:cr&mrp}. To check  that $D'$ is a C-repair of $D$, check for every tuple in $t \in D \smallsetminus D'$, first if  $D \smallsetminus (D' \cup \{t\}) \in
\mc{CT}(D, D,V^\kappa, t)$, which can be done in PTIME. If yes, next check if $t \in  \mc{MRC}(D, V^\kappa)$. The responsibility of $t$ can be computed
by binary search over the set  $\{0\} \cup \{\frac{1}{1 + k}~|~ k = 0, \ldots, n\}$, repeatedly using an algorithm to the {\em Test}: $\rho_{_{\!D\!}}(t) > k$?.
 The cost of the {\em Test} (i.e. the decision problem \nit{RPD}) depends on $\kappa(\mc{Q})$ (as given by
the dichotomy result in \cite{Meliou2010a}).  For each $t$, we need  in the worst case,
essentially $\nit{log(n)}$ calls to the {\em Test}. Considering all tuples, the whole test needs, say a quadratic number of calls to {\em Test}. For weakly-linear queries, this can be done
in polynomial time.

\vspace{1mm}\noindent
(b) There is a repair $D'$  of size greater than $ m >0$  with $t \notin D'$ iff there exists a $t$ and a $\Gamma \subseteq D$, such that
$t$ is an actual cause for $V^\kappa$, and $\Gamma$ is a contingency set for $t$,  $|\Gamma| \leq n-m-1$ and $D' \cap (\{t\} \cup \Gamma) = \emptyset$ iff there is $t$ with $\rho_{_{\!D\!}}(t) >
\frac{1}{n-m}$. So, if the last test is in PTIME, the decision problem about repairs is also in PTIME.

Now, for a given tuple $t$, $\rho_{_{\!D\!}}(t) > \frac{1}{1 + k}$ iff there is a repair $D'$ of $D$ with $t \notin D'$ and $|D'| > n-k -1$.}

\end{document}